\documentstyle[prb,aps,multicol,epsf]{revtex}
\tighten
\narrowtext

\begin{document}
\draft

\title{Level and Eigenfunction Statistics in Billiards with Surface
Scattering}
\author{Ya.~M.~Blanter$^{a,b}$,
A.~D.~Mirlin$^{c,d,*}$, and B.~A.~Muzykantskii$^e$}
\address{
$^a$ D\'epartement de Physique Th\'eorique, Universit\'e de Gen\`eve,
CH-1211 Gen\`eve 4, Switzerland\\
$^b$ Department of Applied Physics and DIMES, Delft University of
Technology, Lorentzweg 1, 2628 CJ Delft, The Netherlands\\
$^c$ Institut f\"ur Nanotechnologie, Forschungszentrum Karlsruhe,
D-76021 Karlsruhe, Germany\\
$^d$ Institut f\"ur Theorie der Kondensierten Materie,
Universit\"at Karlsruhe, D-76128 Karlsruhe, Germany\\
$^e$ Department of Physics, University of Warwick, CV4 7AL Coventry,
UK}
\date{\today}
\maketitle
\begin{abstract}
Statistical properties of billiards with diffusive boundary scattering
are investigated by means of the supersymmetric $\sigma$-model in a
formulation appropriate for chaotic ballistic systems.  
We study level statistics, parametric level statistics, and properties
of electron wavefunctions. In the universal regime, our results
reproduce conclusions of the random matrix theory, while beyond this
regime we obtain a variety of system-specific results determined by
the classical dynamics in the billiard. Most notably,
we find that level correlations do not vanish at arbitrary separation
between energy levels, or if measured at arbitrarily large difference
of magnetic fields. Saturation of the level number variance indicates
strong rigidity of the spectrum. To study spatial correlations of
wavefunction amplitudes, we reanalyze and refine derivation of the
ballistic version of the $\sigma$-model. This allows us to obtain a
proper matching of universal short-scale correlations with
system-specific ones. 
\end{abstract}


\begin{multicols}{2}

\section{Introduction}

Chaotic cavities, commonly understood as quantum systems whose
classical analogues exhibit chaotic dynamics, have become a common
object of research in condensed matter physics. Experimentally,
features of chaotic motion appear, among others, in quantum dots
\cite{Curacao}, microwave cavities \cite{Microwexp}, and small
metallic clusters \cite{clusters1}. For definiteness, we talk below
about electrons in quantum dots (billiards). 

On the theoretical side, properties of chaotic cavities are subdivided
into universal and non-universal. By {\em universal} we mean physical
quantities which only depend on the global symmetry of the system
(like the time reversibility) and possibly trivially scale with the
size of the system, but are not influenced by any cavity-specific
details of dynamics of the electron motion. At the same time, these
quantities differ for systems with chaotic and integrable classical
analogues, and are therefore conceptually important when one discusses
signatures of chaotic behavior in quantum systems. Examples of
universal effects include low-frequency level statistics, leading
order conductance and shot noise. These properties are by now well
understood and described by various types of random matrix theory
(Gaussian ensembles of Hamiltonians for closed or almost closed
systems or circular ensembles of scattering matrices
for open systems).

On the other hand, {\em non-universal} quantities, such as, for
instance, level correlation function for high energy separation
or eigenfunction correlations at large distances, are determined by
sample-specific details of electron motion. These non-universal
quantities thus discriminate between the behavior of individual
billiards. 

A standard tool for treating fluctuations of the density of states
(DOS) in chaotic systems is the real space path integral
approach. Within this method, the DOS correlation function is given by
Gutzwiller's trace formula \cite{Berry,trace1} which has a form of a
sum over periodic orbits of a specific billiard. To evaluate the
formula explicitly in the non-universal regime, one has to resort to
numerical treatment.  

In this paper we will use an alternative approach which has attracted
considerable interest recently, the ballistic $\sigma$-model. It
generalizes the supersymmetric $\sigma$-model, which proved to be very
successful for disordered metals \cite{Efetov,Mirlin}, to ballistic
disordered systems \cite{MK,MK1}. In the framework of this approach, 
all non-universal quantities are expressed through eigenvalues and
eigenfunctions of the Liouville operator, which introduces
non-universality. It has also been conjectured that the same
$\sigma$-model in the limit of vanishing disorder describes
statistical properties of spectra of individual classically chaotic
system.  This conjecture was further developed in
Refs. \onlinecite{Agam,AASA,AASA1} where the $\sigma$-model was 
obtained by means of energy averaging, and the Liouville operator was
replaced by its regularization --- the Perron-Frobenius operator. The
progress along this direction is complicated by the fact that the 
eigenvalues of the Perron-Frobenius operator for many systems are
unknown, while its eigenfunctions can be extremely singular. 

We thus conclude that it is highly desirable to have an example of a
chaotic system which is analytically solvable. We would expect that
universal properties of such an example will conform with the predictions
of the random matrix theory (RMT), whereas explicit expressions for
non-universal quantities would improve our understanding of properties of
chaotic cavities.

Currently, we are unaware of such an example with chaotic
dynamics. However, one can instead treat systems with surface
disorder, which leads to diffusive scattering at the boundary of a
billiard. This model mimics the behavior of a system in the hard chaos
regime: as a result of surface disorder any two arbitrary close
trajectories spread apart after the first collision with the
surface. This must be contrasted with slightly distorted integrable
billiard with the typical spatial scale of this distortion being of
the same order as the size of the
system\cite{Roughb1,Frahm1,Frahm2,Frahm3}. Those systems, termed by
the authors {\em rough billiards}, exhibit slow diffusion over angular
momentum. Systems with surface disorder are also different from
integrable systems with {\em bulk} disorder in the ballistic regime
\cite{AG1,AG2,Fishman}. 

Studies of level and eigenfunction statistics \cite{DEK,BMM} have
shown that, indeed, universal properties of a billiard with
diffusive surface scattering agree with the RMT predictions. At the
same time, non-universal features have been found which 
reflect the classical ballistic dynamics in the billiard.
Subsequently, analytical results for persistent currents
\cite{Samokhin} and transport properties \cite{BS00} of chaotic
cavities have been obtained in the same model. Numerically, billiards
with surface disorder have been studied in connection with the level
\cite{Louis,Louis1} and eigenfunction \cite{Louis,Louis2}
statistics, and magnetoconductance \cite{Louis3}. To this end,
Refs. \onlinecite{Louis,Louis1,Louis2,Louis3} consider a lattice model
with the boundary cites having random energies. A treatment of
Ref. \onlinecite{Apel}, which models surface disorder by cutting off
the boundary cites by confining potential, and eventually proceeds
with numerical evaluation of persistent currents, seems to describe a
similar physical situation.

In this article, we perform a systematic analytical study of a circular
billiard with boundary disorder, based on the $\sigma$-model
approach. The paper is organized as follows.  Section \ref{sigmamodel}
presents the $\sigma$-model for a circular billiard with diffusive
boundary scattering. We subsequently use this approach to derive the
results for level statistics (Section \ref{spectral}), parametric
level statistics describing variation of individual energy levels with
applied magnetic field (Section \ref{parametric}), and eigenfunction
correlations (Section \ref{eigen}). In Section \ref{mixed} we
generalize the problem, imposing the mixed boundary condition, instead
of purely diffusive reflection. This boundary condition enables us to
model a broader class of chaotic systems, where the lowest Lyapunov
exponent is parametrically different form the inverse time of
flight. The obtained results are summarized in Section \ref{conclus},
where we also present a discussion of some open problems. A brief
account of the results of Sections \ref{spectral} and \ref{eigen} has
been previously given in Ref. \onlinecite{BMM}; the level statistics
(Section~\ref{spectral}) were independently studied in
Ref. \onlinecite{DEK}.

\section{Circular billiard with diffusive boundary scattering:
$\sigma$-model approach} \label{sigmamodel}

\subsection{General considerations}

We consider a $2D$ circular billiard of a radius $R$, which is clean
(ballistic) inside, and contains some disorder (to be specified below)
at the boundary. Our starting point is the $\sigma$-model for ballistic
disordered systems \cite{MK,MK1}. The effective action for this model
has the form
\begin{eqnarray} \label{model1}
F[g(\bbox{r}, \bbox{n})] & = & - \frac{\pi\nu}{2} \int d\bbox{r} 
{\rm Str} \left[ i\omega \Lambda \langle g(\bbox{r}) \rangle -
\frac{1}{2\tau(\bbox{r})}
\langle g(\bbox{r}) \rangle^2 \right. \nonumber \\
& - & \left. 2v_F \langle \Lambda T^{-1} \bbox{n} \nabla T
\rangle \right].
\end{eqnarray}
Here the semi-classical Green's function $g(\bbox{r}, \bbox{n})$
integrated over energies is a $4\times 4$ supermatrix which depends on
the coordinate $\bbox{r}$ and direction of the momentum $\bbox{n}$. To
simplify the presentation we consider the case when the time-reversal
symmetry is broken in the quantum problem but is preserved in the
classical one, which can be achieved {\em e.g.} by applying a weak
magnetic field. (See discussion in Section~\ref{plcf}). The angular
brackets denote averaging over $\bbox{n}$: $\langle {\cal O} (\bbox{n})
\rangle = \int d{\bbox n} {\cal O}(\bbox{n})$ with the normalization
$\int d{\bbox n} = 1$, and the supertrace is defined as trace of
boson-boson block minus trace of fermion-fermion one. The
matrix $g$ is constrained by the condition $g(\bbox{r}, \bbox{n})^2 =
1$, and generally can be represented as $g = T \Lambda T^{-1}$, with
the matrix $\Lambda = {\rm diag} (1,1,-1,-1)$ discriminating between
retarded and advanced components of the Green's function.  As usual,
$v_F$ and $\nu=m/(2\pi)$ denote the Fermi velocity and the density of
states at the Fermi surface, respectively; $\tau(\bbox{r})$ is the
(position dependent) elastic scattering time, which originates from
the disorder. We use the units with $\hbar=1$ in the rest of the
paper.  

When all the disorder is at the boundary the scattering time 
$\tau(\bbox{r})$ must be chosen in a way that it is infinite
everywhere except for a thin layer around the boundary. In the
consideration below this term only modifies the boundary condition.

The action (\ref{model1}) differs from that in the $\sigma$-model for
diffusive systems \cite{Efetov} in two respects: First, the Green's
function $g$ is defined in phase space; second, Eq. (\ref{model1}) is
linear in gradients, whereas its diffusive counterpart is of second
order in spatial derivatives. Despite this difference, methods developed
for the calculation of level and eigenfunction statistics in diffusive
systems can be applied here. Indeed, these properties are governed by the
structure of the action in the vicinity of the homogeneous configuration
of the $g$-field\cite{other1}, $g(\bbox{r}, \bbox{n})=\Lambda$ (see Refs.
\onlinecite{Efetov,Mirlin} for review).  In this case the action may be
considerably simplified. Writing $T = 1 - W/2+\ldots$, we find the action
to leading order in $W$,
\begin{equation} \label{model2}
  F_0[W] = \frac{\pi\nu}{2} \int d\bbox{r} d\bbox{n} {\rm Str} \left[
    W_{21} \left( \hat K - i\omega \right) W_{12} \right],
\end{equation}
where the indices $1,2$ refer to the ``advanced-retarded'' decomposition
of $W$, and $\hat K \equiv v_F \bbox{n} \nabla$ is the Liouville
operator. This ``linearized'' action has now the same form as that of a
diffusive system, with the diffusion operator $-D\nabla^2$ replaced by
the Liouville operator. Thus, all the results derived previously from the
linearized action for diffusive systems can be directly used for our
model, provided the eigenvalues and eigenfunctions of the diffusion
operator are replaced by those of the operator $\hat K$.

Since disorder is only present in the close vicinity of the boundary,
we model it by supplementing the Liouville operator $\hat K$ by a boundary
condition. Generally, the boundary condition for an eigenfunction
$\varphi(\bbox{r}, \bbox{n})$ relates its values for outgoing and
incoming particles at a point on the surface,
\begin{eqnarray} \label{bound2}
\varphi(\bbox{r}, \bbox{n}) & = & \left( \int_{(\bbox{N}\bbox{n'}) >
0} \left( \bbox{N} \bbox{n'} \right) B (\bbox{n}, \bbox{n}')
d\bbox{n'} \right)^{-1}  \nonumber \\
& \times & \int_{(\bbox{N}\bbox{n'}) > 0}
\left( \bbox{N} \bbox{n'} \right) B (\bbox{n}, \bbox{n}') \varphi
(\bbox{r}, \bbox{n'}) d\bbox{n'}, \nonumber \\
& & \hspace{4.cm} \left (\bbox{N} \bbox{n} \right) < 0,
\end{eqnarray}
with some kernel $B$. Here the point $\bbox{r}$ is taken at the
surface, $\vert r \vert = R$, and $\bbox{N}$ is an outward normal to
the boundary. The form (\ref{bound2}) ensures that no current flows
through the boundary of the billiard.

The scattering kernel $B$ was intensively studied in the context of the
boundary condition for the distribution function (for review, see Ref.
\onlinecite{Okulov}) and found to be model dependent. Realistic models of
short-range surface randomness (like a boundary narrow layer of
impurities, or a ballistic orifice in a disordered medium) lead to
general boundary conditions of the type (\ref{bound2}), where the kernel
$B$ is a parameterless function of order one.

Following Refs. \onlinecite{Fuchs,Ovchin,Abr}, we approximate the
above, rather complicated, boundary condition by a simpler one,
where an electron reflects diffusively with probability $\alpha$ and
specularly with probability $1 - \alpha$ ($0 \le \alpha \le 1$),
\begin{eqnarray} \label{bound3}
\varphi(\bbox{r}, \bbox{n}) & = & \pi\alpha\int_{(\bbox{N}\bbox{n'}) >
0} \left( \bbox{N} \bbox{n'} \right) \varphi (\bbox{r}, \bbox{n'})
d\bbox{n'} \nonumber \\
& + & (1-\alpha) \varphi (\bbox{r}, \bbox{n''}), \ \ \ \ \ \left
(\bbox{N} \bbox{n} \right) < 0,
\end{eqnarray}
where the vector $\bbox{n''}$ is chosen so that the reflection
from $\bbox{n''}$ to $\bbox{n}$ is purely specular. Although
Eq. (\ref{bound3}) does not correspond to any particular known
microscopic model of disorder, it is commonly believed to provide a
good qualitative description of surface scattering interpolating
between purely diffusive ($\alpha = 1$) and purely specular ($\alpha =
0$) reflection.

With the exception of Section \ref{mixed}, we consider below purely
diffusive reflection (Eq. (\ref{bound3}) with $\alpha = 1$). Physically,
it describes quantum scattering due to short-range disorder (correlated
at the scale of the order of the wavelength $2\pi/p_F$).
Alternatively, the diffusive boundary condition of this type may result
from the purely classical scattering off a strongly corrugated surface.
As noted in the Introduction, atomic-scale disorder has the feature that
an electron loses the memory about the direction of previous motion after
the first collision with the boundary. The system is thus described by
{\em two} characteristic energies, which are the mean level spacing
$\Delta$ and the inverse time of flight through the billiard $v_F/R$.
This is a special type of a billiard, analogous to ``hard chaos''
behavior of genuinely chaotic systems.

Section \ref{mixed} is devoted to the general situation $\alpha < 1$.  A
new regime appears for $\alpha \ll 1$ when the time $R/(v_F\alpha)$
during which an electron remembers its initial direction of motion differs
parametrically from the time of flight $R/v_F$. In this regime the system
is described by {\em three} distinct energy scales.

\subsection{Eigenvalues of the Liouville operator for diffusive
scattering}
\label{eigenvalues}

As we have mentioned, the level statistics of our billiard are entirely
determined by the eigenvalues of the Liouville operator $\hat K$. Due to
the boundary condition~(\ref{bound3}) the evolution becomes irreversible,
the eigenvalues of $\hat K$ have positive real part, and no
regularization, like that discussed in Refs. \onlinecite{AASA,AASA1} is
needed. Below we review properties of eigenvalues $\lambda$ of the
operator $\hat K$,
\begin{equation} \label{eigenv1}
v_F \bbox{n} \nabla \varphi_{\lambda} (\bbox{r}, \bbox{n}) = \lambda
\varphi_{\lambda} (\bbox{r}, \bbox{n}),
\end{equation}
supplemented by the boundary condition
\begin{figure}
\narrowtext
{\epsfxsize=5.83cm\centerline{\epsfbox{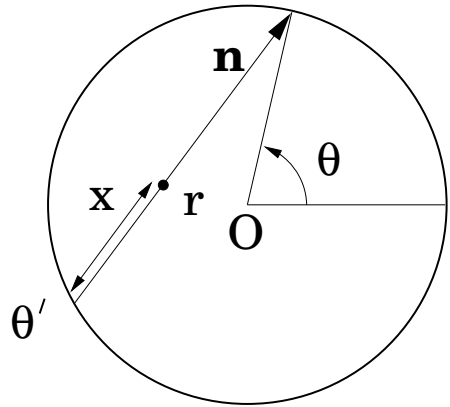}}}
\vspace{0.4cm}
\caption{Natural parameterization of the constant energy shell in
terms of  position $\bbox r = (r, \vartheta)$ and direction of motion
$\bbox n$ and its relation to the alternative  parameterization by the
coordinates $\theta, \theta',x$ introduced in the text.}
\label{fig1}
\end{figure}
\begin{equation} \label{bound1}
\varphi(\bbox{r}, \bbox{n}) = \pi \int_{(\bbox{N}\bbox{n'}) > 0}
\left( \bbox{N} \bbox{n'} \right) \varphi (\bbox{r}, \bbox{n'})
d\bbox{n'}, \ \ \ \left (\bbox{N} \bbox{n} \right) < 0.
\end{equation}

The constant energy surface can be parameterized by the three real
numbers $(\theta, \theta',x)$, where the angle $\theta$ ($\theta'$)
corresponds to the point where the straight line passing through
$\bbox r= (r, \vartheta)$ in the direction $\bbox n$ ($\bbox {-n}$)
crosses the boundary, and $x$ is the distance from the boundary to
$\bbox r$ along this straight line (see Fig.~\ref{fig1}). To cover the
whole energy surface the parameters must change in the region $0 <
\theta,\theta' <2 \pi$; $0<x<2R \sin \vert (\theta - \theta')/2
\vert$. Eq. (\ref{eigenv1}) then expresses the eigenfunction $\varphi$
at any $x$ through the eigenfunction at $x=0$, {\em i.e.}  through the 
function describing particles scattered from the boundary. In
particular, the eigenfunction for the particles arriving to the
boundary is 
\begin{eqnarray} \label{eigenv2}
& & \varphi_{\lambda} \left (\theta, \theta', x = 2R \sin \left\vert
\frac{\theta -
\theta'}{2} \right\vert \right) \nonumber \\
& & = \varphi_{\lambda} \left (\theta, \theta', x = 0 \right) \exp
\left( \frac{2R\lambda}{v_F} \sin \left\vert \frac{\theta -
\theta'}{2} \right\vert \right).
\end{eqnarray}
Now the boundary condition (\ref{bound1}) is used to find a closed
integral equation for $ \varphi_{\lambda} (\theta, \theta', x = 0)$ which
is simplified by the Ansatz $\varphi_{\lambda} (\theta, \theta', x =
0) = \tilde\varphi (\theta - \theta') \exp(il\theta')$, $l$ having the
meaning of the angular momentum,
\begin{equation} \label{eigenv3}
\tilde \varphi (\theta) e^{-il\theta} = \frac{1}{2} \int_0^{\pi}
d\tilde\theta\ \sin \tilde\theta e^{2\xi\sin \tilde\theta}
\tilde\varphi(2\tilde\theta),
\end{equation}
with the notation $\xi \equiv R\lambda/v_F$. Eq. (\ref{eigenv3}) only has
solutions for the eigenvalues which obey the following equation,
\begin{equation} \label{values}
\tilde J_l(\xi) \equiv -1 + \frac{1}{2} \int_0^{\pi} d\theta\ \sin\theta
\exp \left[ 2il\theta + 2 \xi \sin\theta \right]  = 0.
\end{equation}

The eigenvalue equation (\ref{values}) can not be solved analytically
in a closed form, however the combinations of eigenvalues which enter
level statistics can be expressed through the function $\tilde
J_l$. Below we list some properties of these eigenvalues.
\begin{figure}
\narrowtext
{\epsfxsize=7cm\centerline{\epsfbox{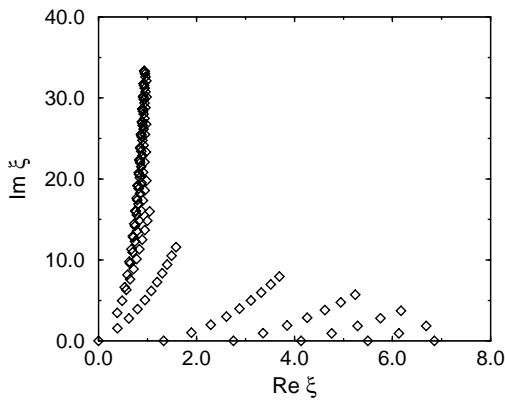}}}
\caption{The first $11\times 11$  ($0 \le k,l < 11$)
  eigenvalues of the Liouville operator $\hat K$ in units of $v_F/R$, as
  given by roots of Eq. (\protect\ref{values}).}
\label{fig2}
\end{figure}

For each value of $l=0,\pm1,\pm2,\ldots$ Eq.(\ref{values}) has a set of
solutions $\xi_{lk}$ with $\xi_{lk}=\xi_{-l,k}=\xi^*_{l,-k}$, which can
be labeled with $k=0,\pm1,\pm2,\ldots$ (even $l$) or
$k=\pm1/2,\pm3/2,\ldots$ (odd $l$). Thus, the eigenvalues form a
two-parameter set. For $l=k=0$ we have $\xi_{00}=0$, corresponding to the
zero mode $\varphi (\bbox{r}, \bbox{n})=\mbox{const}$. All other
eigenvalues have positive real part $\mbox{Re}\,\xi_{lk} > 0$ and govern
the relaxation of the corresponding classical system to the homogeneous
distribution in the phase space.

The asymptotic form of the solutions of Eq.(\ref{values}) for large
$\vert k \vert$ and/or $\vert l \vert$ is given by the saddle-point
method:
\begin{eqnarray} \label{valas}
\xi_{kl} \approx \left\{ \begin{array}{lr} 0.66 l + 0.14 \ln l + 0.55
\pi i k, & 0 \le k \ll l \\
(\ln k)/4 + \pi i (k+1/8), & 0 \le  l \ll k
\end{array} \right. .
\end{eqnarray}
Note that for $k=0$ all eigenvalues are real, while for high values of
$k$ they lie close to the imaginary axis and do not depend on
$l$. Fig.~2 shows a plot of the first 11$\times$11 eigenvalues of the
operator $\hat K$.

\subsection{Green's function of the Liouville operator}

The Green's function of the Liouville operator ${\cal D} (\bbox{r}_1,
\bbox{n}_1; \bbox{r}_2, \bbox{n}_2)$ is the time-integrated
probability to find the particle at the point of the phase space
$(\bbox{r}_1, \bbox{n}_1)$ if it started the motion at $(\bbox{r}_2,
\bbox{n}_2)$. This function obeys the equation
\begin{eqnarray} \label{green0}
& & \hat K {\cal D} (\bbox{r}_1, \bbox{n}_1; \bbox{r}_2, \bbox{n}_2) =
\nonumber \\
& & \qquad \frac{1}{\pi\nu} \left[ \delta(\bbox{r}_1 - \bbox{r}_2)
\delta(\bbox{n}_1 - \bbox{n}_2) - \frac{1}{V} \right],
\end{eqnarray}
where $V = \pi R^2$ is the area of the billiard, and the operator $\hat
K$ acts on the variables $\bbox{r}_1, \bbox{n}_1$. The Green's function
is related to the correlation of the wavefunctions (Section
\ref{eigen}). It is also used in the alternative derivation of the
low-energy level correlation function, see Section \ref{altern}. 

\subsubsection{Green's function integrated over momenta}

In this subsection we calculate the Green's function integrated over
momenta:
\begin{equation} \label{greenint0}
\Pi_B(\bbox{r}_1, \bbox{r}_2) = \int d\bbox{n}_1 d\bbox{n}_2 {\cal D}
(\bbox{r}_1, \bbox{n}_1; \bbox{r}_2, \bbox{n}_2),
\end{equation}
which describes the probability to find an electron at $\bbox{r}_1$
after it has been found at $\bbox{r}_2$, and is important for the
wavefunction correlation. Integrating Eq. (\ref{green0}) over
$d\bbox{n}_2$, we obtain
\begin{equation} \label{greenint1}
v_F \bbox{n}_1 \frac{\partial h}{\partial {\bbox r}_1} = \frac{1}{\pi\nu}
\left[ \delta(\bbox{r}_1 - \bbox{r}_2) - \frac{1}{V} \right],
\end{equation}
where $h(\bbox{r}_1, \bbox{n}_1; \bbox{r}_2) = \int d\bbox{n}_2
{\cal D}$. To solve Eq. (\ref{greenint1}), we use the same strategy as with
Eq. (\ref{eigenv1}) and replace the coordinates $(\bbox{r}_1,
\bbox{n}_1)$ by the variables $\theta_1, \theta_1', x_1$,
\begin{eqnarray} \label{greenint2}
& & h(\theta_1, \theta_1', x_1; \bbox{r}_2) = h(\theta_1, \theta_1',
x_1 = 0; \bbox{r}_2) \nonumber \\
& + & \frac{2}{p_F} \int_0^{x_1} dx_1'' \left[ \delta(\bbox{r}
(\theta_1, \theta_1', x_1'') - \bbox{r}_2) - \frac{1}{V} \right].
\end{eqnarray}
The function $h(\theta_1, \theta_1', x_1=0; \bbox{r}_2)$ which describes
the particles leaving the boundary at the point $(R, \theta_1)$, does not
depend on $\theta_1'$. Using the boundary condition for the operator
$\hat K$, we obtain the equation for this function (below the irrelevant
arguments are dropped and the function is denoted by
$h(\theta_1;\bbox{r_2})$),
\begin{eqnarray} \label{greenint3}
h(\theta_1;\bbox{r_2}) & = & \frac{1}{4} \int_0^{2\pi} d\theta'' \sin
\frac{\theta''}{2} h(\theta_1 + \theta''; \bbox{r_2}) \nonumber \\
& + & \frac{1}{2p_F} \int_{0}^{2\pi}
d\theta'' \int_0^{2R \sin(\vert \theta_1 - \theta'' \vert/2)} dx_1''
\sin \frac{\theta''}{2} \nonumber \\
& \times & \left[ \delta(\bbox{r} (\theta'', \theta_1, x_1'') -
\bbox{r}_2) - \frac{1}{V} \right].
\end{eqnarray}
Calculating the second term in the r.h.s., we obtain
\begin{eqnarray} \label{greenint4}
& & h(\theta_1;\bbox{r_2}) = \frac{1}{4} \int_0^{2\pi} d\theta'' \sin
\frac{\theta''}{2} h(\theta_1 + \theta'';\bbox{r_2}) \\
& & + \frac{1}{p_F} \left[ F(r_2,\theta_1 - \vartheta_2) -
\frac{1}{V} \int \tilde r_2 d\tilde r_2 d \tilde\vartheta_2 
F(\tilde r_2,\theta_1 - \tilde\vartheta_2) \right]. \nonumber 
\end{eqnarray}
Here we have introduced the function
\begin{displaymath}
F(r,\theta) = \frac{R - r \cos \theta}{R^2 + r^2 - 2R r \cos
\theta}, 
\end{displaymath}
and used polar coordinates $(r_k,\vartheta_k)$ for the point $\bbox
r_k$, $k=1,2$. 

Eq. (\ref{greenint4}) is solved by expanding the function
$h(\theta_1;\bbox{r_2})$ in a Fourier series; in this way we find all
the Fourier components $h_k$ except for $k=0$. The component $h_0$ is
not determined by Eq. (\ref{greenint4}) and must be fixed by the
conditions that $\Pi_B(\bbox{r}_1, \bbox{r}_2)$ is symmetric and its
integral over $d\bbox{r}_1$ equals zero (the conservation of the
number of particles). Restoring subsequently the function $h$ in the
bulk of the billiard from $h(\theta_1;\bbox{r_2})$ by means of Eq.
(\ref{greenint2}) and integrating it over $d\bbox{n}_1$, we obtain
the Green's function of the Liouville operator integrated over momenta,
\begin{eqnarray} \label{greenint5}
& & \Pi_B(\bbox{r}_1, \bbox{r}_2) = f_0(\bbox{r}_1, \bbox{r}_2) +
f_1(\bbox{r}_1, \bbox{r}_2), \nonumber \\
& & f_0(\bbox{r}_1, \bbox{r}_2) = \frac{1}{\pi p_F} \left\{
\frac{1}{\vert \bbox{r}_1 - \bbox{r}_2 \vert} - \frac{1}{V} \int
d \bbox {\tilde r}_1 \frac{1}{\vert  \bbox{\tilde r}_1 - \bbox{r}_2
\vert} \right. \nonumber \\
& & - \left. \frac{1}{V} \int d \bbox{\tilde r}_2 \frac{1}{\vert
\bbox{r}_1 - \bbox{\tilde r}_2 \vert} + \frac{1}{V^2} \int d
\bbox{\tilde r}_1 d \bbox{ \tilde r}_2
\frac{1}{\vert \bbox{\tilde r}_1 - \bbox{\tilde r}_2 \vert} \right\}, \\
& & f_1(\bbox{r}_1, \bbox{r}_2) = \frac{1}{4\pi p_F R}
\sum_{k=1}^{\infty} \frac{4k^2 - 1}{4k^2} \left( \frac{r_1 r_2}{R^2}
\right)^k \cos k(\vartheta_1 - \vartheta_2). \nonumber
\end{eqnarray}
It can be actually traced in the course of the calculation that the term
$f_0$ in Eq. (\ref{greenint5}) describes the propagation of an electron
from $\bbox{r}_2$ to $\bbox{r}_1$ along the straight line (no collisions
with the surface). Likewise, the term $f_1$ describes the processes which
involve at least one collision; furthermore, in the factor $4k^2 - 1$ in
$f_1$ the term $4k^2$ originates from the trajectories with one
collision, whereas $-1$ is related to the double and multiple collisions.

\subsubsection{Full Green's function}

Analogously to the previous subsection the value of the function
${\cal D}$ in the bulk of the sample is expressed through its value on
the boundary. Instead of Eq. (\ref{greenint2}) we now have
\begin{eqnarray} \label{green1}
& & {\cal D}(\theta_1, \theta_1', x_1; \bbox{r}_2, \bbox{n}_2) =
{\cal D}(\theta_1, \theta_1', x_1 = 0; \bbox{r}_2, \bbox{n}_2)\nonumber \\
& + & \frac{2}{p_F} \int_0^{x_1} dx_1'' \left[ \delta(\bbox{r}
(\theta_1, \theta_1', x_1'') - \bbox{r}_2) \delta \left(\bbox{n}_1 -
\bbox{n}_2 \right) - \frac{1}{V} \right]. \nonumber\\
&&
\end{eqnarray}
Due to the boundary condition, the function ${\cal D}(\theta_1;
\bbox{r}_2, \bbox{n}_2) \equiv {\cal D}(\theta_1, \theta_1', x_1 = 0;
\bbox{r}_2, \bbox{n}_2)$ does not depend on $\theta_1'$, and obeys the
equation 
\begin{eqnarray} \label{green2}
& & {\cal D}(\theta_1; \bbox{r}_2, \bbox{n}_2) = \frac{1}{4}
\int_0^{2\pi} d\theta'' \sin 
\frac{\theta''}{2} {\cal D}(\theta_1 + \theta''; \bbox{r}_2,
\bbox{n}_2) \nonumber \\ 
& + & \frac{1}{2p_F} \int_0^{2\pi} d\theta'' \int_0^{2R \sin(\vert
\theta_1 - \theta'' \vert/2)} dx_1'' \sin \frac{\theta''}{2} \nonumber
\\
& \times & \left[ \delta(\bbox{r} (\theta'', \theta_1, x_1'')
- \bbox{r}_2) \delta \left(\bbox{n} (\theta'', \theta_1) - \bbox{n}_2
\right) - \frac{1}{V} \right].
\end{eqnarray}
After lengthy calculations, we find that the second term in the r.h.s.
equals $(2\pi/p_FR)[\delta(\theta_1 - \theta_2') - 1/2 \pi]$, where
$\theta_2'$ is the polar angle corresponding to the point at the boundary
to which the vector $\bbox{n}_2$ points from $\bbox{r}_2$. For
definiteness, we restrict all the angles $\theta_1, \theta_1', \theta_2,
\theta_2'$ to the interval $[0, 2\pi]$. Continuing in the same
way as in the previous section, we find the final expression for the
Green's function of the Liouville operator,
\begin{eqnarray} \label{green3}
& & {\cal D} (\bbox{r}_1, \bbox{n}_1; \bbox{r}_2, \bbox{n}_2) = {\cal D}_0 +
{\cal D}_1 + {\cal D}_2, \\
& & {\cal D}_0 = \frac{1}{\pi p_F} \left\{ \delta \left( \frac{\bbox{r}_1 -
\bbox{r}_2}{\vert \bbox{r}_1 - \bbox{r}_2 \vert} - \bbox{n}_1 \right)
\frac{\delta(\bbox{n}_1 - \bbox{n}_2)}{\vert
\bbox{r}_1 - \bbox{r}_2 \vert} \right. \nonumber \\
& & \ \ \ - \left. \frac{1}{V} \int d\bbox{\tilde r}_1\ \delta \left(
\frac{\bbox{\tilde r}_1 - \bbox{r}_2}{\vert \bbox{\tilde r}_1 -
\bbox{r}_2 \vert} 
- \bbox{n}_1 \right) \frac{\delta(\bbox{n}_1 -
\bbox{n}_2)}{\vert\bbox{\tilde r}_1 - \bbox{r}_2 \vert} \right. \nonumber \\
& & \ \ \ - \left. \frac{1}{V} \int d\bbox{\tilde r}_2\ \delta \left(
\frac{\bbox{r}_1 - \bbox{\tilde r}_2}{\vert \bbox{r}_1 - \bbox{\tilde
r}_2 \vert} 
- \bbox{n}_1 \right) \frac{\delta(\bbox{n}_1 - \bbox{n}_2)}{\vert
\bbox{r}_1 - \bbox{\tilde r}_2 \vert} \right. \nonumber \\
& & \ \ \ + \left. \frac{1}{V^2} \int d\bbox{\tilde r}_1 d\bbox{\tilde
r}_2\ \delta 
\left( \frac{\bbox{\tilde r}_1 - \bbox{\tilde r}_2}{\vert \bbox{\tilde
      r}_1 - \bbox{\tilde r}_2
\vert} - \bbox{n}_1 \right) \frac{\delta(\bbox{n}_1 -
\bbox{n}_2)}{\vert \bbox{\tilde r}_1 - \bbox{\tilde r}_2 \vert}
\right\}; \nonumber \\
& & {\cal D}_1 = \frac{2\pi}{Rp_F} \left[ \delta(\theta_1 - \theta_2') -
\frac{1}{2\pi} \right]; \nonumber \\
& & {\cal D}_2 = -\frac{1}{24Rp_F} \left[ 3(\theta_1 - \theta_2')^2 -
6\pi \vert \theta_1 - \theta_2' \vert + 2\pi^2 \right]. \nonumber
\end{eqnarray}
It is straightforward to check that the integration of
Eq. (\ref{green3}) over $d\bbox{n}_1 d\bbox{n}_2$ gives
Eq. (\ref{greenint5}). The Green's function (\ref{green3}) possesses
obvious properties ${\cal D}(\bbox{r}_1, \bbox{n}_1; \bbox{r}_2,
\bbox{n}_2) = {\cal D}(\bbox{r}_2, -\bbox{n}_2; \bbox{r}_1, -\bbox{n}_1)$
and $\int d\bbox{r}_1 d\bbox{n}_1 {\cal D}(\bbox{r}_1, \bbox{n}_1;
\bbox{r}_2, \bbox{n}_2) = \int d\bbox{r}_2 d\bbox{n}_2 {\cal
D}(\bbox{r}_1, \bbox{n}_1; \bbox{r}_2, \bbox{n}_2) = 0$.

Actually, ${\cal D}_0$ is a solution of Eq. (\ref{green0}) in the
infinite space and thus represents propagation processes which do not
involve any collisions with the boundary. This kind of propagation is
only possible if $\bbox{n}_1$ coincides with $\bbox{n}_2$ and both are
directed along $\bbox{r}_1 - \bbox{r}_2$. Furthermore, ${\cal D}_1$ is
responsible for the propagation from $\bbox{r}_2$ to $\bbox{r}_1$ with
only one intermediate collision, which necessarily requires $\theta_1
= \theta_2'$. Finally, the term ${\cal D}_2$ describes propagation
which involves two or more intermediate collisions.

\section{Spectral statistics} \label{spectral}

In this Section we discuss the level correlation function and level
number variance of the circular billiard with the boundary condition
(\ref{bound1}).

As we have mentioned in the Introduction, the results for the level
statistics follow readily from general formulae derived for the diffusive
conductors\cite{Mirlin} where the eigenvalues of the diffusion operator
are replaced by those of the Liouville operator.

\subsection{Low frequencies}

We define the level correlation function in a standard way,
$$R_2(\omega) = (V\Delta)^{2} \langle \nu(\epsilon + \omega)
\nu(\epsilon) \rangle-1,$$
where $\nu(\epsilon)$ is the (fluctuating) density of states, $\Delta
=(V \nu)^{-1}$ is the mean level spacing.

In the range of relatively low frequencies (which in our case
means $\omega \ll v_F/R$, see below) the function $R_2(\omega)$ quite
generally has the form \cite{KM} ($s = \omega/\Delta$)
\begin{eqnarray} \label{lowen}
R_2(s) &=& \delta(s) - {\sin^2 \pi s\over (\pi s)^2} +
\frac{A}{\pi^2 g_b^2} \sin^2\pi s,\\
g_b & =& \frac{v_F}{R\Delta}=\frac{p_FR}{2}.\nonumber
\end{eqnarray}
The first two terms are universal and actually correspond to the
random matrix theory result. They are associated with the contribution
of the mode with zero eigenvalue of the Liouville
operator\cite{Efetov}. The last term represents the non-universal
correction; the information about the operator $\hat K$ enters through
the dimensionless constant $A=\sum'\xi_{kl}^{-2}$, where the prime
indicates that the eigenvalue $\xi_{00}=0$ is excluded.  The value of
$A$, as well as the high-frequency behavior of $R_2(s)$ (see below),
can be extracted from the spectral function \cite{ASh}
\begin{equation} \label{sum1}
S (\omega) = \sum_{l} S_l (\omega);\ \ \ S_l(\omega) \equiv \sum_k
\left( \lambda_{kl} - i\omega \right)^{-2}.
\end{equation}
According to the Cauchy theorem, $S_l$ can be represented as an
integral in the complex plane,
\begin{equation} \label{Cauchy1}
S_l (\omega) = \left( \frac{R}{v_F} \right)^2 \frac{1}{2\pi i} \oint_C
\frac{1}{(z - i\omega R/v_F)^2} \frac{\tilde J_l' (z)}{\tilde J_l (z)}
dz,
\end{equation}
where the contour $C$ encloses all zeroes of the function $\tilde J_l
(z)$. Evaluating the residue at $z=i\omega R/v_F$, we find
\begin{equation} \label{sl}
S_l (\omega) = - (R/v_F)^2 \left. \frac{d^2}{d z^2} \right\vert_{z =
  i\omega R/v_F} \ln \tilde J_l (z) .
\end{equation}
Considering the limit $\omega \to 0$ and subtracting the contribution
of $\lambda_{00}=0$, after lengthy but elementary algebra we obtain
\begin{eqnarray} \label{s0}
 A & = & \left( \frac{v_F}{R} \right)^2 S(0) \nonumber \\
& = & -19/27 - 175 \pi^2/1152 + 64/(9\pi^2) \approx -1.48.
\end{eqnarray}
In contrast to the diffusive case \cite{KM}, this constant is negative:
the level repulsion is {\em enhanced} with respect to result for RMT.  We
recollect that in the diffusive case the level repulsion is {\em
  suppressed} as compared to RMT, and this suppression has a plausible
physical explanation. Indeed, the non-universal correction in diffusive
case is proportional to $g_d^{-2}$, where $g_d = E_c/\Delta \gg 1$ is the
dimensionless conductance, and $E_c$ is the Thouless energy. In
disordered conductors, the parameter $g_d$ is responsible for the
metal-insulator transition/crossover. This non-universal term thus
reflects a tendency to localization in level statistics \cite{KM} and
would become of the order of unity for $g_d \sim 1$, when the system
approaches the insulating regime (with uncorrelated levels, $R_2 (s) =
\delta(s))$.

In contrast, the last term of Eq. (\ref{lowen}) describes different
physics. For ballistic systems, the limit $g_b \sim 1$ means not the
insulating regime, but rather a quantum-mechanical system far from the
semi-classical regime. Energy levels in these systems may be strongly
correlated, even more strongly than in RMT, hence the tendency to
enhancement of the level repulsion observable from
Eq. (\ref{lowen}). We will see that this tendency is even more
pronounced in the high-energy behavior of the level correlation
function.  

The last point we make is the following. Eq.(\ref{lowen}) is valid as
long as the correction is small compared to the RMT result,
i.e. provided $\omega$ is below the inverse time of flight,
$v_F/R$. Thus, the inverse time of flight serves as a Thouless energy
for this ballistic billiard, with $g_b$ playing the role of the
dimensionless conductance. Note that $g_b$ is related to the number
$N$ of levels below the Fermi energy as $g_b=N^{1/2}$.

\subsection{Low frequencies -- Alternative derivation} \label{altern}

The constant $A$ can also be found in another way, using the Green's
function of the Liouville operator. Indeed, the quantity $S(0)$ can be
written in the following form,
\begin{eqnarray} \label{greenstat1}
S(0) & = & (\pi\nu)^2 \int d\bbox{r}_1 d\bbox{r}_2 d\bbox{n}_1
d\bbox{n}_2 {\cal D}(\bbox{r}_1, \bbox{n}_1 ; \bbox{r}_2, \bbox{n}_2)
\nonumber \\
& \times & {\cal D}(\bbox{r}_2, \bbox{n}_2; \bbox{r}_1, \bbox{n}_1),
\end{eqnarray}
where the function ${\cal D}$ is given by Eq. (\ref{green3}).
Eq. (\ref{greenstat1}) is directly verified by expanding the Green's
function in the eigenfunctions of the operator $\hat K$. Now we
calculate the integral in Eq. (\ref{greenstat1}) directly. It is
instructive to split the result into three terms,
\begin{eqnarray} \label{greenstat2}
S(0) & = & S_A + S_B + S_C, \nonumber \\
S_A & = & \int {\cal D}_0 {\cal D}_0 = \left( \frac{R}{v_F} \right)^2
\left( -1 + \frac{64}{9\pi^2} \right); \nonumber \\ 
S_B & = & \int ({\cal D}_0{\cal D}_1 + {\cal D}_1{\cal D}_0 + {\cal
D}_1{\cal D}_1) \nonumber \\ 
& = & \left( \frac{R}{v_F}
\right)^2 \left( -\frac{8}{3} + \frac{\pi^2}{8} \right); \nonumber \\
S_C & = & \int ({\cal D}_0{\cal D}_2 + {\cal D}_2{\cal D}_0 + {\cal
D}_1{\cal D}_2 + {\cal D}_2{\cal D}_1 + 
{\cal D}_2{\cal D}_2) \nonumber \\
& = & \left
( \frac{R}{v_F} \right)^2 \left( -\frac{80}{27} - \frac{2\pi^2}{9} -
\frac{\pi^2}{16} + \frac{2\pi^2}{128} \right),
\end{eqnarray}
which reproduces Eq. (\ref{s0}). In our notation, $S_A$ is given by the
processes which involve no collisions with the boundary, $S_B$ is a
contribution of trajectories with only one collision, and $S_C$ takes
into account all other processes. We see that all these contributions are
of the same order and can not be disregarded.

\subsection{High frequencies}

In the range $\omega\gg\Delta$ the level correlation function can be
decomposed into the smooth (Altshuler-Shklovskii) part\cite{ASh},
\begin{equation}
\label{as}
R_2^{\mbox{\small AS}} (\omega) = (\Delta^2/ 2\pi^2) {\rm
Re}\,S(\omega),
\end{equation}
and the part $R_2^{\mbox{\small osc}}$
which oscillates on the scale of the level spacing\cite{AAn}. We
consider below the high-frequency regime $\omega \gg v_F/R$.

\subsubsection{Smooth part}

For high frequencies, the integral in the definition (\ref{values}) of
the functions $\tilde J_l$ is small, and thus $\tilde J_l$ is close to
$-1$. Expanding the logarithm in Eq. (\ref{sl}) up to second order in
$\tilde J_l + 1$ and using the Poisson formula,
\begin{equation} \label{Poisson1}
\sum_l e^{2il\theta} = \pi \sum_{n=-\infty}^{\infty} \delta(\theta -
\pi n),
\end{equation}
to sum over $l$, we obtain for $\omega \gg v_F/R$
\begin{equation} \label{highen0}
R_2^{\mbox{\small AS}} (\omega) = \frac{1}{g_b^2} \left(
\frac{v_F}{2\pi\omega R} \right)^{1/2} \cos \left( 4 \frac{\omega
R}{v_F} - \frac{\pi}{4} \right). 
\end{equation}

Thus, the smooth part of the level correlation function is an
oscillating function of frequency, with the period of $\pi v_F/2R$ and
a slowly decaying amplitude. To clarify the connection with the
periodic orbit theory, we note that any closed sequence of chords
joined at vertices is a ``periodic orbit'' in our model. The above
period corresponds to the ``periodic orbit'' which traverses twice the
diameter of the billiard (the longest ``periodic orbit'' with two
vertices). The amplitude of the oscillations is proportional to
$g_b^{-2}$. This is very different from the diffusive regime, where
the smooth part of the level correlation function does not exhibit any
oscillations. Furthermore, in 2D case the AS contribution vanishes,
and the leading behavior is provided by weak localization effects
\cite{KL}. 

\subsubsection{Oscillating part}

The oscillating part of the level correlation function
$R_2^{\mbox{osc}}(s)$ for frequencies $\omega \gg \Delta$ is given by
\cite{AAn}
\begin{equation} \label{highen}
R_2^{\mbox{osc}} (s)  = (1/2\pi^2)\cos(2\pi s) D(s),
\end{equation}
where $D(s)$ is the spectral determinant,
\begin{equation} \label{spectr11}
D(s) = s^{-2}\prod_{kl\ne(00)}  (1- is\Delta/\lambda_{kl} )^{-1}
(1 + is\Delta/\lambda_{kl} )^{-1}.
\end{equation}

Since $\Delta^{-2}\partial^2\ln D(s)/\partial
s^2=-2\mbox{Re}\,S(\omega)$, the spectral determinant can be recovered
from Eq. (\ref{sl}),
\begin{equation} \label{spdet1}
D(s) = \left( \frac{\pi}{2} \right)^6 \frac{e^{c_1 + c_2s}}{g_b^2}
\prod_l \frac{1}{\tilde J_l (i s /g_b) \tilde J_l (-i s /g_b)},
\end{equation}
where $c_1$ and $c_2$ are arbitrary constants which are fixed by the
requirement that Eq.(\ref{highen}) in the range $\Delta\ll\omega\ll
v_F/R$ reproduces the low-frequency behavior (\ref{lowen}).  Expanding
$\ln D$ in $s/g_b \ll 1$ and comparing it with the low-frequency
expression
$$\ln D(s) = \frac{1}{s^2} - \frac{A}{g_b^2},$$
which follows from Eq.(\ref{lowen}), we obtain $c_1 = c_2 =
0$. Finally, for $s \gg 1$ ($\omega \gg v_F/R$) Eq. (\ref{spdet1})
yields for the oscillating part of the
level correlation function,
\begin{equation} \label{highen2}
R_2^{\mbox{osc}} (\omega) = \frac{\pi^4}{128g_b^2}
\cos \left( \frac{2\pi \omega}{\Delta} \right).
\end{equation}
It is remarkable that the amplitude of the oscillating part does not
depend on frequency. This is in contrast with the diffusive case, where
in the AS regime ($\omega$ above the Thouless energy) the oscillating
part $R_2^{\mbox{osc}} (\omega)$ is exponentially small \cite{AAn}.  Such
a behavior indicates a strong rigidity of a level system and is
reminiscent of 1D harmonic oscillator.

\subsection{Spectral form-factor}

\begin{figure}
\narrowtext
\epsfxsize=7cm\centerline{\epsfbox{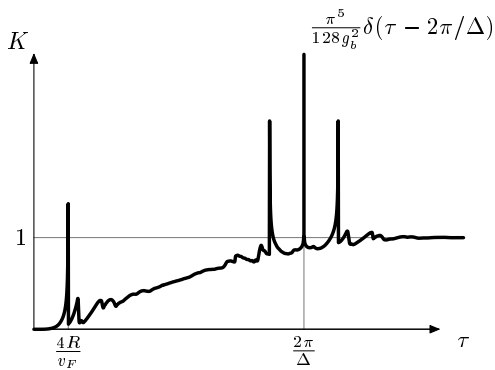}}
\caption[short caption]{
  \label{k(t)}
  Form factor for the two point correlation function $K(\tau)=\int d
  \omega e^{i\omega\tau} R_2(\omega)$ for $g_b=5$. The
  $\delta$-function at $\tau= 2 
  \pi/\Delta$ is shown as a vertical spike. It arises because of the
  non-decaying oscillations (\ref{highen2}).  The $1/\sqrt{T_2-\tau}$
  singularity at $T_2=4R/v_F$ is related to the ``periodic orbit'' which
  has period $T_2$ and traverses twice the diameter of the billiard. The
  same orbit gives rise to singularities at $\tau=2 \pi /\Delta \pm
  T_2$}
\end{figure}

To illustrate the nature of the oscillating terms in the asymptotes
(\ref{highen0}), (\ref{highen2}) we compute numerically the form factor $
K(\tau)=\int d \omega e^{i\omega\tau} R_2(\omega) $ of the two-level
correlation function for $g_b=5$. The result is shown on Fig.~\ref{k(t)}.
The non-zero limit at $\omega \to \infty$ of the oscillating part
$R_2^{\mbox{osc}}(\omega)$ gives rise to a $\delta$-function contribution
of the form $\pi^5/(128 g_b^2) \delta(\tau- 2 \pi /\Delta)$ (shown as a
vertical spike on Fig.~\ref{k(t)})\cite{lastfoot}. The ``periodic
orbit'' which traverses the billiard twice along the diameter and has
the period $T_2=4R/v_F$ manifests itself as a set of power-law
singularities in $K(\tau)$ at $\tau=T_2$ and $\tau= 2\pi/\Delta \pm
T_2$.  Near to these singularities the form factor diverges like
$1/\sqrt{T_2-\tau}$ and $1/\sqrt{T_2 \pm (2\pi/\Delta -\tau)}$,
respectively. The peak to the right of the first singularity is the
contribution from the ``equilateral triangle'' periodic orbit with the
period $T_3 = 3 \sqrt{3} R/v_F$. The same orbit causes small peaks at
$\tau = 2 \pi/\Delta \pm T_3$.

\subsection{Level number variance}

In practice, the description of level statistics by the means of the
level correlation function is not always convenient. Indeed, for low
frequencies $\omega \ll v_F/R$ non-universal effects are manifested
only as small corrections to RMT. For high frequencies, the entire
behavior is non-universal, but in this range the level correlation
function is proportional to $g_b^{-2}$ and small. Therefore, in order
to study non-universal effects in level statistics, experimentally or
by means of computer simulations, it is instructive to consider 
a quantity more sensitive to these effects.

A well-known way to emphasize the high-energy behavior of the level
correlation function is to study the variance of the number of levels
in an energy interval of width $E = s\Delta$,
\begin{equation} \label{lnv}
\Sigma_2 (s) = \int_{-s}^s \left(s - \vert \tilde{s} \vert
\right) R_2(\tilde{s}) d\tilde{s},
\end{equation}
Direct calculation gives for $ s \ll g_b$ ($E \ll v_F/R$)
\begin{equation}
  \pi^2 \Sigma_2 (s)
  =  1 + \gamma + \ln (2\pi s) +
    \frac{A s^2}{2g_b^2} \label{lnv1}
\end{equation}
and for $s \gg g_b$ ($E \gg v_F/R$)
\begin{eqnarray}
  \pi^2 \Sigma_2 (s) &=& 1 + \gamma + \ln \frac{16 g_b}{\pi^2}
  \nonumber \\
  &&  - \frac{\pi^2}{16}  \left( \frac{2 g_b}{\pi s}
  \right)^{1/2} \cos \left( \frac{4 s}{g_b} - \frac{\pi}{4}
  \right).
\label{lnv2}
\end{eqnarray}
Here $\gamma \approx 0.577$ is Euler's constant, and $A$ is defined by
Eq.(\ref{s0}). The first three terms at the r.h.s. of Eq.(\ref{lnv1})
represent the RMT contribution, which is logarithmic in energy
\cite{Mehta}. It is remarkable that the actual level number variance
is always {\em smaller} than that given by RMT.
\begin{figure}
\narrowtext
{\epsfxsize=7cm\centerline{\epsfbox{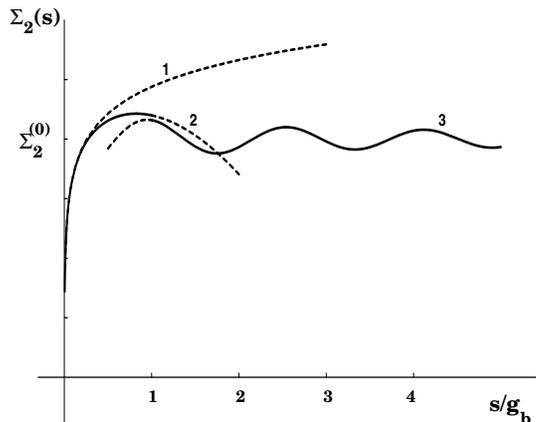}}}
\caption{Level number variance $\Sigma_2 (E)$ as a function of
energy; $s = E/\Delta$. Curve 1 shows the RMT result, while curves 2
and 3 correspond to asymptotic regimes of low (\protect\ref{lnv1}) and
high (\protect\ref{lnv2}) frequencies. The saturation value
$\Sigma_2^{(0)}$ is given by Eq. (\protect\ref{sigmasat}).}
\label{fig3}
\end{figure}

As seen from Fig.~\ref{fig3}, the two asymptotes (\ref{lnv1}) and
(\ref{lnv2}) perfectly match in the intermediate regime, $s\sim g_b$
($E \sim v_F/R$). Taken together, they provide a complete description
of $\Sigma_2(s)$. According to Eq.(\ref{lnv2}), the level number
variance saturates at the value
\begin{equation} \label{sigmasat}
\Sigma_2^{(0)} = \pi^{-2} (1 + \gamma + \ln (16 g_b/\pi^2)) \gg 1.
\end{equation}
This saturation again confirms the conclusion that we have already
made from analyzing the low-energy behavior of the level correlation
function --- the system of levels of our billiard is quite rigid, more
rigid than in RMT. We also remark that a saturation of $\Sigma_2 (s)$,
as well as its oscillations on the scale set by the shortest periodic
orbit, is predicted by Berry for a generic chaotic billiard
\cite{Berry,Berry1}. This behavior is also in agreement with the
results for $\Sigma_2 (s)$ found numerically for a tight-binding model
with moderately strong disorder on boundary sites
\cite{Louis,Louis1,foot1}.

At the same time, we note that the behavior of $\Sigma_2 (s)$ is quite
different from that in other reference systems. Indeed, for both
diffusive systems \cite{ASh} and systems with bulk disorder in
ballistic regime \cite{AG1,AG2,Fishman} the level number variance is
{\em greater} than RMT. The energy at which $\Sigma_2$ is expected to
saturate depends on the type of disorder. For short-range impurities
(white noise random potential) arbitrarily short periodic orbits
exist, and thus no saturation up to $\omega \sim E_F$ is expected. For
a diffusive system and smooth random potential the diffusive dynamics
lead to a linear increase of the level number variance,
$\Sigma_2(s)\sim s/g_d$, for $s\gg g_d\sim \nu v_Fl_{tr}$, while the
shortest orbits have lengths of the order of the transport mean free
path $l_{tr}$, causing the saturation of $\Sigma_2 (s)$ at a
parametrically larger value  $s \sim v_F/(l_{tr}\Delta)$. 
Similarly, in ``rough billiards'' (slightly distorted
integrable billiards \cite{Frahm1,Frahm3}) the level number variance
is higher than in RMT and does not saturate until a value 
$s\sim v_F/R\Delta$ parametrically exceeding the effective Thouless
energy, since the system is diffusive in the angular momentum space.

\section{Parametric level statistics} \label{parametric}

\subsection{Introduction}

In this Section, we study the parametric statistics of energy
levels of our system. Specifically, we assume that the billiard
is placed in a magnetic field, which plays the role of an
external parameter.

Already a considerable amount is known on the
subject (see Ref. \onlinecite{GMW} for review). In this paper
we are interested in the parametric level correlation function
\begin{eqnarray} \label{param0}
& & R_{\Phi} (\omega, B) = -1 \\
& + & (V\Delta)^2 \left\langle \nu(E + \omega/2, \bar B +
B/2) \nu(E - \omega/2, \bar B - B/2) \right\rangle, \nonumber
\end{eqnarray}
where the mean magnetic field $\bar B$ is introduced in order to
break time-reversal symmetry. The correlation function
(\ref{param0}) has been previously investigated in the
diagrammatic expansion \cite{Szafer}, the $\sigma$-model approach
\cite{SA,AAn}, the random matrix theory (by means of Dyson's brownian
motion model) \cite{Been1,Been2}, and by semi-classical methods
\cite{Goldberg,Ozorio}. The most remarkable observation of these works
is that when the perturbation is weak {\em and} the frequency $\omega$
is low, the parametric correlation function is universal, {\em i.e} it
does not depend on the type of the perturbation, provided it is
properly rescaled, nor on any details of the system.

Below we study {\em non-universal} behavior of the parametric
level correlation function  (\ref{param0}). We use the
$\sigma$-model formulation of parametric statistics developed in
Ref.  \onlinecite{SA} to derive an analog of
Altshuler-Shklovskii formula for this function and evaluate
it explicitly for our model.

\subsection{Eigenvalues of the Liouville operator in magnetic
field}

The correlation function (\ref{param0}) may be obtained from the
supersymmetric $\sigma$-model with the effective action (\ref{model1})
provided the operator $\nabla$ is replaced by the ``gauge-invariant''
combination $\nabla - (ie/2c)\Lambda \bbox{A}$, where $\bbox{A}$ is the
vector potential which corresponds to the field $B$: $B = \nabla \times
\bbox{A}$. Repeating the steps leading to Eq. (\ref{model2}) we find that
in the effective action the Liouville operator $\hat K$ is replaced by
its gauge-invariant form in the magnetic field,
\begin{equation} \label{gauge1}
\hat K_{\Phi} = v_F \bbox{n} \left( \nabla - \frac{ie}{c} \bbox{A}
\right).
\end{equation}

Prior to the investigation of statistical properties of the energy
levels, we find the eigenvalues of the operator $\hat
K_{\Phi}$. Choosing the symmetric gauge $\bbox{A} = \bbox{B} \times
\bbox{r}/2$ we obtain, instead of Eq.~(\ref{values}), the equation 
\begin{eqnarray} \label{values1} 
& & J^{\phi}_l(\xi) \equiv -1 \\
& + & \frac{1}{2} \int_0^{\pi} d\theta\ \sin\theta
\exp \left[ 2il\theta + 2 \xi \sin\theta + i \phi \sin 2\theta \right]
= 0, \nonumber
\end{eqnarray}
which determines the dependence of the eigenvalue $\lambda_l = \xi
v_F/R$ on the magnetic field. We have introduced the dimensionless
parameter $\phi = \Phi/\Phi_0$, where $\Phi = \pi BR^2$ is the
magnetic flux through the billiard, and $\Phi_0 = 2\pi\hbar c/e$ is
the flux quantum. 

Equation (\ref{values1}) defines a two-parameter set of eigenvalues
$\xi_{kl}$. It stays invariant under the simultaneous transformations
$l \to -l$, $\xi \to \xi^*$, $\phi \to -\phi$, which means that if
$\lambda (B)$ is an eigenvalue of the operator $\hat K_{\Phi}$ then
$\lambda^* (-B)$ is also an eigenvalue.

For sufficiently low frequencies and magnetic fields the
parametric correlation function (\ref{param0}) is dominated by
the zero mode $l=k=0$ (see below). To find the evolution of the
corresponding eigenvalue $\xi_{00}$, we put $l = 0$ and take a
variation of Eq. (\ref{values1}) with respect to $\phi \ll 1$,
\begin{eqnarray*}
\int_0^{\pi} d\theta\ \sin\theta \left[ 2 \xi_{00} \sin\theta +
i \phi \sin 2\theta - \frac{\phi^2}{2} \sin^2 2\theta \right] =
0,
\end{eqnarray*}
which gives $\xi_{00} = 8\phi^2/(15\pi)$.

\subsection{Parametric level correlation function}
\label{plcf}

The parametric correlation function $R_{\Phi}$ (\ref{param0}) is
expressed in terms of the retarded $G^R$ and advanced $G^A$ Green's
functions in the following way,
\begin{eqnarray} \label{param1}
R_{\Phi} (\omega, B) & = & \frac{1}{2} \left[ T(\omega, B) + T(-\omega,
-B) \right], \\
T(\omega, B) &\equiv& \frac{1}{2(\pi\nu)^2} \left\langle \mbox{\rm Tr}
\ G^R (E+\omega/2, \bar B + B/2) \right. \nonumber \\
& \times & \left. \mbox{\rm Tr} \ G^A (E-\omega/2, \bar
B -B/2) \right\rangle_c, \nonumber
\end{eqnarray}
where $\langle AB \rangle_c \equiv \langle AB \rangle - \langle A
\rangle \langle B \rangle$ denotes the irreducible part (cumulant).

For low frequencies {\em and} magnetic fields (the precise condition is
specified below) the function $R_{\Phi}$ is non-perturbative and can be
found using the random matrix theory. This regime was previously
investigated in Ref.~\onlinecite{SA}. Here we focus instead on the case
of higher frequencies and/or fields, when the smooth part of the
parametric correlation function $R^{AS}_{\Phi} (\omega, B)$ is
correctly described by perturbation theory. From
Eq. (\ref{param1}) we obtain 
\begin{eqnarray} \label{param2}
& & R^{AS}_{\Phi} (\omega, B) \nonumber \\
& = & \frac{\Delta^2}{4\pi^2} \mbox{\rm Re}
\sum_{kl} \left\{ \frac{1}{[-i\omega + \lambda_{kl} (B)]^2} +
\frac{1}{[i\omega + \lambda_{kl} (-B)]^2} \right\} \nonumber \\
& = & \frac{\Delta^2}{2\pi^2} \mbox{\rm Re}
\sum_{kl} \frac{1}{[-i\omega + \lambda_{kl} (B)]^2}.
\end{eqnarray}

\begin{figure}
\narrowtext
\epsfxsize=7cm\centerline{\epsfbox{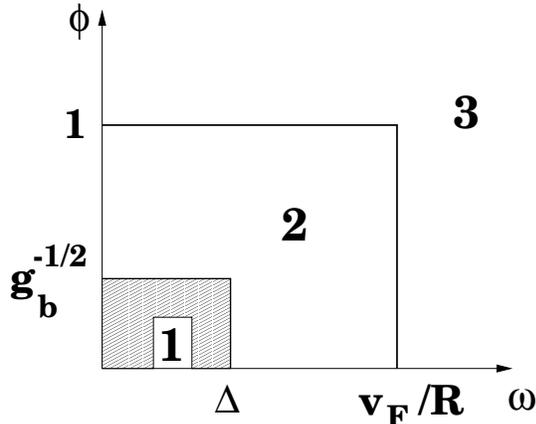}}
\caption{Different parameter regions (in the frequency -- magnetic field
  plane) for the parametric level statistics: 1 -- non-perturbative (RMT)
  region, 2 -- zero-mode region, 3 -- Altshuler-Shklovskii region}
\label{f:parametic}
\end{figure}

To identify relevant parameters, we consider $R^{AS}_{\Phi}$ at zero
frequency. For low fields, the sum in Eq. (\ref{param2}) is dominated by
the lowest eigenmode, which as we have seen is quadratic in field,
$\lambda_{00} \sim v_F\phi^2/R$. The perturbation theory does not apply
for $\lambda_{00} \lesssim \Delta$, which corresponds to the fields $\phi
\lesssim g_b^{-1/2}$, where as before $g_b = v_F/(R\Delta)$. Generally, a
non-perturbative calculation is needed when $\omega \lesssim \Delta$ 
{\em and} $\phi \lesssim g_b^{-1/2}$ (region 1 in
Fig.~\ref{f:parametic}). Everywhere outside this regime, the
perturbative expression (\ref{param2}) applies. 

For higher fields, the zero mode still dominates until $\phi
\sim 1$. At $\phi \sim 1$ other modes become important, and, in
addition, the zero-mode eigenvalue can not be taken quadratic in
field any more. Thus, the point $\phi \sim 1$ plays for
parametric correlations the same role as $\omega \sim v_F/R$ for the
usual (non-parametric) level correlation function. Generally,
for $\phi \ll 1$ and $\omega \ll v_F/R$ we are in the regime when
everything is determined by the zero-mode approximation. For
higher fields $\phi \gg 1$ {\em or} frequencies $\omega \gg v_F/R$,
the system crosses over to the Altshuler-Shklovskii regime (region 3 in
Fig.~5), when all the modes are important.

To end our qualitative discussion, we determine the strength of the
magnetic field needed to strongly affect the classical dynamics. The
cyclotron radius of an electron trajectory in the magnetic field is
$r_c = mv_Fc/(eB)$. The magnetic field strongly affects the
dynamics provided $r_c \lesssim R$, which gives $\phi \gtrsim
g_b$. Our theory is thus valid for $\phi \ll g_b$, which still leaves
a large window for the regime $\phi \gg 1$.

\subsubsection{Perturbative zero-mode regime (Region 2 in Fig.~5)}

Using the result for the zero-mode eigenvalue, we find from
Eq. (\ref{param2})
\begin{equation} \label{param4}
R^{AS}_{\Phi} (s, \phi) = \frac{225}{2} \frac{64g_b^2 \phi^4 -
225\pi^2 s^2}{(64g_b^2 \phi^4 + 225\pi^2 s^2)^2},
\end{equation}
where, as previously, $s = \omega/\Delta$. The result (\ref{param4})
applies when both $\phi \ll 1$ and $s \ll g_b$, but either $\phi \gg
g_b^{-1/2}$ or $s \gg 1$. Without the magnetic field, $\phi = 0$ and we
reproduce $R^{AS}_{\Phi} (s,0) = -(2\pi^2s^2)^{-1}$, which is a smooth
version of the RMT result (\ref{lowen}). For $s=0$, we obtain
$$R^{AS}_{\Phi} (0, \phi) = \frac{225}{128 g_b^2 \phi^4}.$$

\subsubsection{Altshuler-Shklovskii regime (Region 3 in Fig.~5)}

For $\phi \gg 1$ or $s \gg g_b$ we generalize Eq. (\ref{sl}) to the
case of parametric statistics,
\begin{equation} \label{sl1}
R^{AS}_{\Phi} (s, \phi) = - \frac{1}{2\pi^2g_b^2} \mbox{\rm Re} \sum_l 
\left. \frac{d^2}{d z^2} \right\vert_{z = i s/g_b} \ln J_l^{\phi}
(z),
\end{equation}
and use the fact that $J_l^{\phi} \simeq -1$. Expanding $\ln
J_l^{\phi} (z)$ in $J_l^{\phi} + 1$ up to the second order, we
obtain
\end{multicols}
\widetext
\hrulefill\hfill\
\begin{eqnarray} \label{sl2}
R^{AS}_{\Phi} (s, \phi) & = & - \frac{1}{2\pi^2g_b^2} \mbox{\rm Re}
\left. \sum_l \left\{ -2 \int_0^{\pi} d\theta \ \sin \theta
\exp(2il\theta + 2z \sin \theta + i\phi \sin 2\theta) \right.
\right. \nonumber \\
& - & \Biggl. \Biggl. \frac{1}{8} \frac{d^2}{d z^2} \int_0^{\pi}
d\theta_1 d\theta_2 \sin\theta_1 \sin\theta_2 \exp \left[
2il(\theta_1 + \theta_2) + 2z (\sin \theta_1 + \sin \theta_2) +
i\phi (\sin 2\theta_1 + \sin 2\theta_2) \right] \Biggr\}
\Biggr\vert_{z = i s/g_b} .
\end{eqnarray}
\hfill\hrulefill
\begin{multicols}{2}

Now we employ the Poisson formula (\ref{Poisson1}). The first
integral in Eq. (\ref{sl2}) vanishes identically, and in
the second one we have $\theta_1 + \theta_2 = \pi$. Under this
condition, the integral {\em does not depend} on the magnetic
field. We thus have
\begin{eqnarray} \label{corparam1}
R^{AS}_{\Phi} (s, \phi) & = & \frac{1}{\pi g_b^2} \mbox{\rm Re}
\int_0^{\pi} d\theta \ \sin^4 \theta \exp(4is \sin\theta /g_b) \\
& \approx & \left\{ \matrix{ (8g_b^2)^{-1} (3 - 20 s^2/g_b^2), &
s \ll g_b
\cr (2\pi s g_b^3)^{-1/2} \cos (4s/g_b - \pi/4), & s \gg g_b
} \right. . \nonumber
\end{eqnarray}
The asymptotes in the regime $s \gg g_b$ are identical to
Eq. (\ref{highen0}).

The result that the parametric level correlation function does not decay
with magnetic field is very different from the diffusive case
\cite{AAn} and is actually quite surprising.  It means that the
density of states at the same energy retains the memory even at very
large fields, after the levels underwent many avoided crossings. The
field dependence of the correlator appears if we keep higher orders of
the expansion of the logarithm in Eq. (\ref{sl1}), but it is only
manifested as small corrections to the result (\ref{corparam1}).

\subsection{Parametric level number variance}

The high-frequency behavior of the non-parametric level
correlation function $R_2 (s)$ is best seen in the level number
variance $\Sigma_2 (s)$, see Section \ref{spectral}. Similarly,
it is instructive to introduce the {\em parametric level number
variance} (PLNV)\cite{Goldberg},
\begin{equation} \label{plnv1}
U(E, B) = \left\langle \left[ N(E,\bar B) - N(E, \bar B + B)
\right]^2 \right\rangle,
\end{equation}
where $N(E, B)$ is the full number of levels in the energy strip $E$
in the magnetic field $B$. In zero field PLNV vanishes (in low fields
it is generally linear in field, see Ref. \onlinecite{SA}). If
the values of $N(E, B)$ and $N(E, \bar B + B)$ are uncorrelated
in high fields (which is not the case in our situation, since
the level correlation function does not decay in high
fields), PLNV saturates, $U(E, B \to \infty) \to 2\Sigma_2 (E)$.

PLNV can be easily expressed through the level correlation
function (analogously to Eq. (\ref{lnv})),
\begin{equation} \label{plnv2}
U(s,\phi) = 2 \int_{-s}^s (s - \vert \tilde s \vert) \left[
R_2(\tilde s) - R_{\Phi} (\tilde s, \phi) \right] d\tilde s,
\end{equation}
where $s = E/\Delta$. Below we concentrate on the perturbative
regime $\phi \gg g_b^{-1/2}$, when
\begin{eqnarray} \label{intermed1}
& & \int_{-s}^s (s - \vert \tilde s \vert) R_{\Phi} (\tilde s, \phi)
d\tilde s \nonumber \\
& \approx & \frac{1}{\pi^2} \mbox{\rm Re} \sum_l \left[ \ln J_l^{\phi}
(is/g_b) - \ln J_l^{\phi} (0) \right]. 
\end{eqnarray}
A direct calculation gives for $g_b^{-1/2} \ll \phi \ll 1$
\begin{eqnarray} \label{plnv3}
U(s, \phi) & = & \frac{2}{\pi^2} (\ln 2\pi s + 1 + \gamma) -
\frac{225}{128} \frac{s^2}{g_b^2 \phi^4}, \nonumber \\
& & \ \ \ \ \ \ \ \ \ \ \ 1 \ll s \ll g_b\phi^2, \\
U(s, \phi) & = & \frac{2}{\pi^2} \left[ \ln \frac{8g_b \phi^2}{15} + 1
+ \gamma \right], \ \ \ g_b\phi^2 \ll s \ll g_b, \nonumber \\
U(s, \phi) & = & \frac{2}{\pi^2} \left[ \ln \frac{32g_b \phi^2}{15} +
1 + \gamma \right] - A', \ \ \ s \gg g_b, \nonumber
\end{eqnarray}
where $\gamma$ is again Euler's constant, and
$$A' = -\frac{2}{\pi^2} \sum_{l=1}^{\infty} \ln \left( 1 -
\frac{1}{4l^2 -1} \right) \approx 0.115.$$

In higher fields $\phi \gg 1$ we have
\begin{eqnarray} \label{pnlv2}
U(s, \phi) & = & \frac{2}{\pi^2} (\ln 2\pi s + 1 + \gamma) +
\frac{s^2}{g_b^2} \left( \frac{A}{\pi^2} - \frac{3}{2} \right),
\nonumber \\
& & \ \ \ \ \ \ \ \ \ \ \ 1 \ll s \ll g_b, \\
U(s, \phi) & = & \frac{2}{\pi^2} \left[ \ln \frac{16g_b}{\pi^2} + 1
+ \gamma \right] - \frac{1}{8}, \ \ \ s \gg g_b,
\nonumber
\end{eqnarray}
where the constant $A$ is given by Eq. (\ref{s0}). A comparison
of Eq. (\ref{pnlv2}) with Eqs. (\ref{lnv1}) and (\ref{lnv2})
shows that in the limit $\phi \to \infty$ PLNV differs from
$2\Sigma_2 (s)$. This difference originates from the fact that
the correlation function $R_{\Phi}$ does not vanish at large
$\phi$ (see Eq. (\ref{corparam1})).

\section{Correlations of eigenfunctions} \label{eigen}

According to Berry's conjecture \cite{Berry2}, a wave function in a 2D
chaotic system shows Gaussian fluctuations with the correlation function
$V\langle\psi^*({\bbox r})\psi({\bbox r'})\rangle=J_0(p_Fr)$ (see
Ref.\onlinecite{Srednicki} for a recent generalization of this
conjecture). The supersymmetry method allows one to derive this result
(which is equivalent to the zero-mode approximation for the
$\sigma$-model) and to calculate system-specific corrections
\cite{Prigodin,FM95,BM97,Prigodin98,Mirlin} .

Let us consider the two-point correlation function
\begin{equation}
C(\bbox{r}_1, \bbox{r}_2) = \Delta\left\langle\sum_\alpha
|\psi_\alpha^2(\bbox{r}_1) \psi_\alpha^2(\bbox{r}_2)|\delta(\epsilon -
\epsilon_\alpha)\right\rangle\ .
\label{double1}
\end{equation}
In the standard diffusive $\sigma$-model \cite{Mirlin}, the correlation
function (\ref{double1}) can be written in the form of the integral
over the $\sigma$-model field Q(\bbox{r}),
\begin{eqnarray}
 C(\bbox{r}_1,\bbox{r}_2)
& = &  \lim_{\eta\to 0}{\eta\Delta\over\pi}
\langle G_{11}(\bbox{r}_1,\bbox{r}_1)G_{22}(\bbox{r}_2,\bbox{r}_2)
\nonumber \\
& + & G_{12}(\bbox{r}_1,\bbox{r}_2)G_{21}(\bbox{r}_2,\bbox{r}_1)
\rangle_{{\cal S}[Q]}\ ,
\label{double3}
\end{eqnarray}
where $\eta=-i\omega/2>0$ is the level broadening, $G$ is the Green's
function in the field $Q$, and the subscripts $1,2$ refer to the
advanced-retarded decomposition, the boson-boson components being
always implied (we drop the corresponding
indices). Eq.~(\ref{double3}) yields the following result \cite{BM97},
\begin{eqnarray}
&& V^2C(\bbox{r}_1,\bbox{r}_2) \simeq  1 + \Pi_D(\bbox{r}_1,\bbox{r}_2) +
k_q(\bbox{r}_1 - \bbox{r}_2)  \nonumber \\
&& + \Pi_D\left({\bbox{r}_1+\bbox{r}_2\over 2},
{\bbox{r}_1+\bbox{r}_2\over 2}\right)
k_q(\bbox{r}_1 - \bbox{r}_2), 
\label{double15}
\end{eqnarray}
where $\Pi_D$ is the diffusion propagator, $k_q(r)=J_0^2(p_Fr)e^{-r/l}$,
and $l$ is the mean free path.  In the framework of the
ballistic $\sigma$-model the diffusive propagator
$\Pi_D(\bbox{r}_1,\bbox{r}_2)$ is replaced by its ballistic counterpart
$\Pi_B(\bbox{r}_1, \bbox{r}_2)$ given in our model by
Eq.~(\ref{greenint5}). At this point we seem to encounter a problem.
Indeed, at short distances $r=|\bbox{r}_1-\bbox{r}_2|$ the classical
propagator $\Pi_B(\bbox{r}_1, \bbox{r}_2)$ is dominated by the
contribution of the direct path,
\begin{equation}
\label{wf1}
\Pi_B(\bbox{r}_1, \bbox{r}_2)\simeq f_0(\bbox{r}_1, \bbox{r}_2) = {1\over
\pi p_F|\bbox{r}_1-\bbox{r}_2|}\ .
\end{equation}
This contribution, which becomes of order unity  at $r\sim\lambda_F$,
would imply strong deviations from the universal Gaussian
statistics. However, it is not difficult to realize that this
contribution is nothing else but a classical ``copy'' of the term
$k_q(\bbox{r}_1-\bbox{r}_2)$. Therefore, we encounter a problem of the
double counting: one and the same contribution is taken into account
twice, classically and quantum-mechanically. In the case of the
diffusive $\sigma$-model this problem does not appear because of the
scale separation: the classical propagator $\Pi_D$ is restricted to 
low momenta $q < l^{-1}$, while the short-scale
($r < l$) physics corresponding to high momenta ($q > l^{-1}$)
is treated quantum-mechanically. 
 
The situation is different in ballistic case.  The semi-classical
description extends now to all momenta $q\lesssim p_F$. Therefore, there
is no separation in momentum space between the slow modes (treated within
the semi-classical approximation) and the fast ones (treated exactly,
{\it i.e.}  quantum-mechanically), and a careful treatment is required in
order to avoid the double counting. For this purpose, we find it
instructive to consider a problem with an additional smooth random
potential $U({\bbox r})$ in the bulk, characterized by a correlation
function $V({\bbox r}-{\bbox r'})=\langle U({\bbox r})U({\bbox
  r'})\rangle$ with a correlation length $d \gg \lambda_F$ and
inducing a small-angle scattering.  Specifically, we will assume that
the corresponding transport scattering rate $\tau_{tr}^{-1}$ is
negligible, {\it i.e.} the transport mean free path
$l_{tr}=v_F\tau_{tr}$ is large, $l_{tr}\gg R$, so that the bulk 
scattering has essentially no effect on the classical dynamics. On the
other hand, the single-particle mean free path $l=v_F\tau$ (corresponding
to the total relaxation rate $\tau^{-1}$) will be assumed to satisfy
$\lambda_F\ll d \ll l \ll R$ and will play a role similar to that of
$l_{tr}$ for the diffusive $\sigma$-model, separating the regions of
classical and quantum treatment. The condition $d \ll l$ ensures that
the scattering on this random potential is of quantum-mechanical
(rather than classical) nature and is correctly treated within the
Born approximation. 

We first ignore the boundary scattering and will return to it in the end
of the calculation.  Averaging over the realizations of the smooth random
potential $U({\bbox r})$, one can derive the $\sigma$-model following
Ref.~\onlinecite{woelfle84}. (This derivation is outlined in Appendix).

To calculate deviations of the eigenfunction statistics from
universality, we write \cite{Mirlin,KM,FM95,BM97} (See Eq. (\ref{wf5}))
\begin{equation}
\label{wf7}
T({\bbox r},{\bbox n}) = T_0(1-W({\bbox r},{\bbox n})/2)\ ,
\end{equation}
and then integrate out perturbatively non-zero modes described by
$W({\bbox r},{\bbox n})$. 
The part of action (\ref{wf5a}) which is quadratic in $W$ has the 
following form in the momentum space,
\end{multicols}
\widetext
\hrulefill\hfill\
\begin{eqnarray}
\label{wf8}
F_0 &=& {1\over 2} {\rm Str} \int d{\bbox n}_1 d{\bbox n}_2 (d{\bbox q}) 
W_{21}(-{\bbox q},{\bbox n}_1) W_{12}({\bbox q},{\bbox n}_2) \nonumber \\
&\times& \left(-{1\over 2} \int(d{\bbox p}) w({\bbox n}_1,{\bbox n})
w({\bbox n},{\bbox n}_2) G_R({\bbox p}_+) G_A({\bbox p}_-)
+ \pi\nu w({\bbox n}_1,{\bbox n}_2) \right) \nonumber\\
&\equiv& {1\over 2}{\rm Str}\int d{\bbox n}_1 d{\bbox n}_2 (d{\bbox q}) 
W_{21}(-{\bbox q},{\bbox n}_1) W_{12}({\bbox q},{\bbox n}_2)
[-A({\bbox q};{\bbox n}_1,{\bbox n}_2) +B({\bbox n}_1,{\bbox n}_2)]\ ,
\end{eqnarray}
\hfill\hrulefill
\begin{multicols}{2} \noindent
where ${\bbox p}_\pm={\bbox p}\pm {\bbox q}/2$. In the last line of
(\ref{wf8}) we introduced the definitions $A$ and $B$ for
the two terms of the quadratic form. Eq.~(\ref{wf8}) induces the
contraction rules for integrals over non-zero modes $W$ ({\em cf.}
Ref. \onlinecite{AL97}), 
\begin{eqnarray}
&& \Bigl\langle \rm{Str}[W(-{\bbox q},{\bbox n}_1)P] 
{\rm Str}[W({\bbox q},{\bbox n}_2)R] \Bigr\rangle \nonumber\\
&& = 2{\cal D}({\bbox q};{\bbox n}_2,{\bbox n}_1) \rm{Str}
P{1-\Lambda\over 2} R{1+\Lambda\over 2} \nonumber \\ && +
2{\cal D}(-{\bbox q};{\bbox n}_1,{\bbox n}_2) \rm{Str} P{1+\Lambda\over 2}
 R{1-\Lambda\over 2}\ , \nonumber \\
&& \Bigl\langle \rm{Str}W(-{\bbox q},{\bbox n}_1)P W({\bbox q},{\bbox
n}_2)R  
\Bigr\rangle \nonumber\\
&& = 2{\cal D}({\bbox q};{\bbox n}_2,{\bbox n}_1) \rm{Str}
P{1+\Lambda\over 2}\  
{\rm Str}  R{1-\Lambda\over 2} \nonumber \\ && +
2{\cal D}(-{\bbox q};{\bbox n}_1,{\bbox n}_2) \rm{Str} P{1-\Lambda\over 2}\ 
\rm{Str} R{1+\Lambda\over 2}\ , \label{wf8b} 
\end{eqnarray}
where $P$ and $R$ are arbitrary matrices and  the propagator ${\cal
D}$ is given by the series
\begin{equation}
\label{wf9}
{\cal D} = (-A+B)^{-1} = B^{-1} + 
B^{-1} A B^{-1} + \ldots\ .
\end{equation}

We are now ready to evaluate the ballistic counterpart of
Eq.~(\ref{double3})  (with the $\sigma$-model field being
$Q(\bbox{r},\bbox{n})$). The Green's function $G$ there is given by
\begin{equation}
\label{wf10}
G = \left(E-H_0-{i\over 2}\int d{\bbox n'}Q({\bbox r},{\bbox n'})
w({\bbox n},{\bbox n'})\right)^{-1} 
\end{equation}
Expanding the term $\langle G_{11} G_{22}\rangle$ in
Eq. (\ref{double3}) up to quadratic order in $W$ and switching to the  
momentum space, we get 
\begin{eqnarray}
\label{wf11}
&& \int d({\bbox r}_1-{\bbox r}_2) \langle G_{11}({\bbox r}_1,{\bbox r}_1)
G_{22}({\bbox r}_2,{\bbox r}_2)\rangle e^{i{\bbox q}({\bbox
r}_1-{\bbox r}_2)} \nonumber\\
&&=-{1\over 4} 
\int (d{\bbox p}_1) d{\bbox n}_1'w({\bbox n}_1,{\bbox n}_1')
 (d{\bbox p}_2) d{\bbox n}_2'w({\bbox n}_2,{\bbox n}_2') \nonumber \\
&&\times \Bigl\langle[G_0({\bbox p}_{1-})Q_1(-{\bbox q},{\bbox n}_1')
G_0({\bbox p}_{1+})]_{11} \Bigr.\nonumber \\
&&\times \Bigl.
[G_0({\bbox p}_{2+})Q_1({\bbox q},{\bbox n}_2')G_0({\bbox p}_{2-})]_{22}
\Bigr\rangle\ , 
\end{eqnarray}
where $G_0=i{\rm Im}G_R\cdot Q_0+{\rm Re}G_R$, 
$Q_0=T_0\Lambda T_0^{-1}$, and 
$Q_1({\bbox q},{\bbox n})=T_0\Lambda W({\bbox q},{\bbox n})T_0^{-1}$. 
Applying the contraction rules (\ref{wf8b}) and then
performing the zero-mode integration, we
reduce the r.h.s. of Eq.~(\ref{wf11}) to the form
\begin{eqnarray}
\label{wf11a}
&&{\Delta\over 4\pi\eta}
\int (d{\bbox p}_1) d{\bbox n}_1' w({\bbox n}_1,{\bbox n}_1')
(d{\bbox p}_2) d{\bbox n}_2'w({\bbox n}_2,{\bbox n}_2') \nonumber \\
&& \times G_R({\bbox p}_{1+})G_A({\bbox p}_{1-})
G_R({\bbox p}_{2+})G_A({\bbox p}_{2-})
{\cal D}({\bbox q};{\bbox n}_2',{\bbox n}_1') \nonumber \\
&& = {\Delta\over \pi\eta w_0^2} \int d{\bbox n}_1'' d{\bbox n}_1'
d{\bbox n}_2'' d{\bbox n}_2' A({\bbox q};{\bbox n}_2'',{\bbox n}_2')
\nonumber\\ 
&&\times {\cal D}({\bbox q};{\bbox n}_2',{\bbox n}_1')
A({\bbox q};{\bbox n}_1',{\bbox n}_1'')\ .
\end{eqnarray}
Note that in order to simplify this expression, we have used 
time reversal invariance of the classical motion, 
${\cal D}(-{\bbox q};{\bbox n},{\bbox n'})= {\cal D}({\bbox q};-{\bbox
n}',-{\bbox n})$. 
Substituting (\ref{wf11a}) in (\ref{double3}), we find
the contribution of the $\langle G_{11}G_{22}\rangle$ term to the wave
function correlator,
\begin{eqnarray}
\label{wf13}
&& V^2C({\bbox r}_1,{\bbox r}_2)|_{\langle G_{11}G_{22}\rangle} -1 
\nonumber\\  && \ \  = {1\over (\pi\nu)^2 w_0^2}
\int d{\bbox n}_1 d{\bbox n}_2 \langle {\bbox n}_1|A{\cal D} A|{\bbox
n}_2\rangle  
\nonumber \\ && \ \  = {1\over (\pi\nu)^2 w_0^2}
\int d{\bbox n}_1 d{\bbox n}_2 \nonumber \\
&& \ \ \times \langle {\bbox n}_1|AB^{-1}A + AB^{-1}AB^{-1}A + 
\ldots  |{\bbox n}_2\rangle\ .
\end{eqnarray}
The r.h.s. of Eq. (\ref{wf13}) is the sum of the ladder diagrams
(corresponding to $1,2,3,\ldots$ intermediate scattering processes)
yielding precisely the classical propagator $\Pi_B({\bbox r}_1,{\bbox
r}_2)$. We see, however, that the first term in this sum
(corresponding to a motion without intermediate scattering) is
absent. This term is equal to 
\begin{eqnarray*}
&& {1\over (\pi\nu)^2 w_0^2} \int d{\bbox n}_1 d{\bbox n}_2 
A({\bbox n}_1,{\bbox n}_2) \\ && = {1\over 2 (\pi\nu)^2}
\int(d{\bbox p})G_R({\bbox  p}_+)G_A({\bbox p}_-)\ ,
\end{eqnarray*}
or, in coordinate space,
\begin{eqnarray*}
 {1\over 2 (\pi\nu)^2}G_R({\bbox r}_1-{\bbox r}_2)
G_A({\bbox r}_2-{\bbox r}_1) &=& 
{e^{-|{\bbox r}_1-{\bbox r}_2|/l} \over \pi p_F|{\bbox r}_1-{\bbox r}_2|} \\
& \equiv & f_0(|{\bbox r}_1-{\bbox r}_2|)\ .
\end{eqnarray*}
Including now the leading contribution of the $\langle
G_{12}G_{21}\rangle$ term in (\ref{double3}), we get the wave function
correlator up to the terms linear in the classical propagator 
$f_0(|{\bbox r}_1-{\bbox r}_2|)$ or in the quantum propagator 
$k_q({\bbox r}_1-{\bbox r}_2)=J_0^2(p_F|{\bbox r}_1-{\bbox r}_2|)
e^{-|{\bbox r}_1-{\bbox r}_2|/l}$,
\begin{equation}
\label{wf14}
V^2C({\bbox r}_1,{\bbox r}_2)\simeq 1 + f_1({\bbox r}_1,{\bbox r}_2) + 
k_q({\bbox r}_1-{\bbox r}_2)\ ,
\end{equation}
where $f_1 = \Pi_B - f_0$. We see that the double counting problem does
not exist anymore.  The quantum ($k_q$) and the classical ($f_1$)
contributions perfectly complement each other, describing the motion
before and after the first collision respectively.

Until now, we did not consider the ballistic version of the last term in
Eq.~(\ref{double15}). Such a term originates from the first order (in
${\cal D}$) correction to $\langle G_{12}G_{21}\rangle$.  For $|{\bbox
r}_1-{\bbox r}_2|\gg\lambda_F$ the corresponding contribution to the
wave function correlator is much smaller than the term coming from
$\langle G_{11}G_{22}\rangle$ (second term in Eq.~(\ref{wf14})), which
is of order of $\Pi_B$, and thus can be neglected.  However, this
contribution becomes important at $|{\bbox r}_1-{\bbox
r}_2|\sim\lambda_F$. In particular, for ${\bbox r}_1={\bbox r}_2$, the 
order-${\cal D}$ contribution from $\langle G_{12}G_{21}\rangle$ is
found to be equal to that from $\langle G_{11}G_{22}\rangle$, yielding 
\begin{equation}
\label{wf16}
V^2C({\bbox r},{\bbox r})\simeq 2 + 2 f_1({\bbox r},{\bbox r})\ .
\end{equation}
After integration over ${\bbox r}$ Eq.~(\ref{wf16}) determines
non-universal correction to the average inverse participation ratio
\begin{equation}
V \left\langle \int d{\bbox r} |\psi({\bbox r})|^4 \right\rangle=2 +
\frac{1}{8\pi g_b} \left[ \ln g_b + O(1) \right]. 
\end{equation}

In the discussion above we did not take into account the boundary
scattering which determines the classical propagator $\Pi_B({\bbox
r}_1,{\bbox r}_2)$ on the scale of the system size $R$ but is 
irrelevant for the matching of the classical and quantum contributions
on the short scale $l \ll R$. In principle, one could avoid
introducing the additional smooth potential and consider the boundary
randomness only. By analogy with the above consideration, we expect that
in this case the classical propagator $f_1$ entering Eq.~(\ref{wf14})
will describe the motion starting from the first collision with the
boundary.

\section{Generalization to mixed boundary condition} \label{mixed}

In the case of the mixed boundary condition (\ref{bound3}) the trace of
the resolvent ${\rm Tr}(\hat{K}-i\omega)^{-1}$ acquires cuts in addition
to simple poles which were present for purely diffusive scattering,
$\alpha=1$ (see the discussion below). For this reason, we cannot write
the spectral function $S(\omega)$ in the Altshuler-Shklovskii form
(\ref{sum1}) but have to use a more general representation,
\begin{eqnarray}
\label{mixed1}
S(\omega) & = &  {\partial^2\over\partial\omega^2} {\rm Tr}\ln
(\hat{K}-i\omega) \nonumber \\
& = &  {\partial\over\partial(i\omega)} \int_0^\infty dt e^{i\omega t}
{\rm Tr} e^{-\hat{K}t} \nonumber \\
& = &  {\partial\over\partial(i\omega)} \int_0^\infty dt e^{i\omega t}
\int d\bbox{r} d\bbox{n} g(\bbox{r}, \bbox{n}; \bbox{r}, \bbox{n}; t)\
,
\end{eqnarray}
where $g(\bbox{r}_1, \bbox{n}_1; \bbox{r}_2, \bbox{n}_2; t)$ is the
{\em time-dependent} Green's function of the operator $\hat{K}$
characterizing the probability of propagation from the point
$(\bbox{r}_2, \bbox{n}_2)$ of the phase space to the point
$(\bbox{r}_1, \bbox{n}_1)$ in a time $t$.

Further transformations are straightforward though somewhat
lengthy. The trace of the Green's function in Eq. (\ref{mixed1}) can be
written as a sum of the terms with $n=2,3,\ldots$ boundary scattering
events. Changing the variables from $(\bbox{r}, \bbox{n})$ to
$(\theta,\theta',x)$ (see Sec.~\ref{eigenvalues}) and performing the
integration over $x$, we can present the result in the form
\begin{eqnarray}
\label{mixed2}
{\rm Tr}(\hat{K}-i\omega)^{-1} &=& {2R\over v_F}\int d\theta d\theta'
\left|\sin {\theta -\theta' \over 2}\right| \nonumber \\
&\times& \sum_{n=2}^\infty
\hat{\Phi}^{n-1}(\theta,\theta';\theta,\theta')\ ,
\end{eqnarray}
where $\hat{\Phi}$ is the integral operator,
\begin{displaymath}
\left[ \hat\Phi f \right] (\theta_1, \theta_1') = \int_0^{2\pi}
d\theta_2 d\theta_2'  \Phi (\theta_1, \theta_1'; \theta_2, \theta_2')
f(\theta_2, \theta_2'),
\end{displaymath}
with the kernel
\begin{eqnarray}
\label{mixed3}
&& \Phi(\theta_1,\theta_1';\theta_2,\theta_2') =
\delta(\theta_1-\theta_2')\exp\left\{ {2i\omega R\over v_F}
\left|\sin {\theta_2-\theta_2' \over 2}\right|\right\} \nonumber\\
&& \times
\left\{(1-\alpha)\delta(\theta_1-\theta_1'-\theta_2+\theta_2') +
{\alpha\over 4}
\left|\sin {\theta_2-\theta_2' \over 2}\right|\right\}\ .
\end{eqnarray}
Physically, $\hat{\Phi}$ characterizes the probability of the scattering
process when a particle moving along the segment $(\theta_2\to\theta_2')$
is reflected into the segment $(\theta_1\to\theta_1')$. Resumming the
series (\ref{mixed2}) and using that, according to Eq. (\ref{mixed3}),
\begin{equation}
\label{mixed4}
{\partial \Phi (\theta_1,\theta_1';\theta_2,\theta_2') \over
\partial (i\omega)} = {2R\over v_F}
\left|\sin {\theta_2-\theta_2' \over 2}\right|
\Phi (\theta_1,\theta_1';\theta_2,\theta_2')\ ,
\end{equation}
we rewrite Eq.~(\ref{mixed2}) in a compact form
\begin{equation}
\label{mixed5}
 {\rm Tr}\ln (\hat{K}-i\omega) = {\rm Tr} \ln (1-\hat{\Phi})\ ,
\end{equation}
up to an irrelevant additive constant (which we drop below).

Eigenfunctions of $\hat{\Phi}$ have the form
\begin{equation}
\label{mixed6}
f(\theta,\theta')=e^{il\theta'} g(\theta-\theta')\ ,
\end{equation}
so we can write
\begin{equation}
\label{mixed7}
{\rm Tr}\ln (\hat{K}-i\omega) = \sum_l {\rm Tr} \ln (1-\hat{\Phi}_l)\ ,
\end{equation}
where $\hat{\Phi}_l$ is obtained by restricting the operator
$\hat{\Phi}$  to the space spanned by functions (\ref{mixed6}) with
a particular $l$.  The kernel of $\hat{\Phi}_l$  is given by
\begin{eqnarray}
\label{mixed8}
\Phi_l(\theta,\theta') &=& \exp\left\{il\theta' + 2i {\omega R\over v_F}
|\sin(\theta'/2)|\right\} \nonumber \\
& \times & \left\{(1-\alpha)\delta(\theta-\theta') + {\alpha\over 4}
|\sin (\theta'/2)|\right\}\ ,
\end{eqnarray}
where $\theta=\theta_1-\theta_1'$, $\theta'=\theta_2-\theta_2'$. We
further represent $\Phi_l$ as a sum of the contributions corresponding
to the two terms in the curly brackets in Eq. (\ref{mixed8}),
\begin{equation}
\label{mixed9}
\Phi_l(\theta,\theta') = \Phi_l^{\rm spec}(\theta')
\delta(\theta-\theta')  + \Phi_l^{\rm diff}(\theta')\ .
\end{equation}
Here $\Phi_l^{\rm spec}$ is associated with the processes of specular
reflection, while $\Phi_l^{\rm diff}$ corresponds to the diffusive
scattering processes. Using Eq. (\ref{mixed9}),  we expand ${\rm
Tr}\ln(1-\hat{\Phi}_l)$ in powers of $\Phi_l^{\rm diff}$ and then
re-sum the series, using the structure of the two terms in
Eq. (\ref{mixed9}) (the first one is proportional to
$\delta(\theta-\theta')$, while the second one independent of
$\theta'$) to get 
\begin{eqnarray} 
\label{mixed10}
& & {\rm Tr}\ln(1-\hat{\Phi}_l) = \int_0^{2\pi} d\theta \ln
(1-\Phi_l^{\rm spec}(\theta)) \nonumber \\
& + & \ln \left[ 1+ \int_0^{2\pi} d\theta (1-\Phi_l^{\rm
spec}(\theta))^{-1} \Phi_l^{\rm diff}(\theta)\right]\ .
\end{eqnarray}
Combining Eqs. (\ref{mixed7}), (\ref{mixed8}), and (\ref{mixed10}) we
finally obtain the following representation for the spectral determinant
of the Liouville operator,
\begin{eqnarray}
\label{mixed11}
& & {\rm Tr}\ln (\hat{K}-i\omega) = \sum_l \left\{ \int_0^\pi d\theta
\right. \\
& & \qquad \times \left. \ln \left[1-(1-\alpha)e^{2i(\omega
R/v_F)\sin\theta + 2il\theta}\right] \right. \nonumber \\
& + & \left. \ln \left[1- {\alpha\over 2}\int_0^\pi d\theta
{\sin\theta\over e^{-2i(\omega R/v_F)\sin\theta - 2il\theta}-1+\alpha}
\right]\right\}\ . \nonumber
\end{eqnarray}

The first term in Eq. (\ref{mixed11}) originates from the first term
in the r.h.s.  of Eq. (\ref{mixed10}) and is determined only by 
$\Phi_l^{\rm spec}$ and not by $\Phi_l^{\rm diff}$. It thus knows only
about the motion in a clean system (without boundary scattering) and
about the total probability $\alpha$ to be scattered away, but not
about the differential scattering probability. In other words, this
term would describe the correlations of the density of states if the
electrons simply disappear (get absorbed) at the boundary with
probability $\alpha$, otherwise being reflected specularly. It
characterizes thus the spectrum of a clean circle (with energy levels
broadened due to the absorbing boundary). Due to these correlations,
the {\em disorder averaged} density of states $\langle \nu (\epsilon)
\rangle$ {\em at given energy} $\epsilon$ fluctuates.  These
fluctuations are not of interest here; one can get rid of them by
subtracting the (energy-dependent) disconnected part of the level
correlation function, thus modifying the definition of $R_2 (\omega)$,
\begin{equation}
\label{mixed12}
R_2^d (\omega)= (V\Delta)^2
[\langle\nu(\epsilon+\omega)\nu(\epsilon)\rangle
-\langle\nu(\epsilon+\omega)\rangle\langle\nu(\epsilon)\rangle]\ ,
\end{equation}
see Ref.~\onlinecite{AG2} for an extensive discussion. We mention also
that, in fact, the first term in Eq. (\ref{mixed11}) does not yield these
correlations fully correctly, since the $\sigma$-model is not appropriate
for the description of integrable systems. Agam and Fishman \cite{Fishman}
used Berry-Tabor trace formula to calculate the analogous contribution of
trajectories not scattered by impurities in an integrable system with
bulk disorder. We do not enter a more detailed discussion here since we
are not interested in this contribution anyway.

The second term in Eq.~(\ref{mixed11}) describes the disorder-induced
correlations. Before turning to the calculation of the level correlation
function, we analyze the structure of singularities of the resolvent
${\rm Tr}(\hat{K}-i\omega)^{-1}$. As in the $\alpha=1$ case, it has
simple poles lying in the half-plane ${\rm Im}\ \omega<0$ which positions
$\xi=i\omega R/v_F$ are determined by the equation
\begin{equation}
\label{mixed13}
{\alpha\over 2}\int_0^\pi d\theta {\sin\theta\over
e^{-2il\theta-2\xi\sin\theta} - (1-\alpha)} = 1\ .
\end{equation}
However,  in contrast to the $\alpha=1$ case, the resolvent
additionally now has branch cuts induced by zero values of the
denominator in Eq. (\ref{mixed13}). These branch cuts can be
parameterized as
\begin{equation}
\label{mixed14}
\xi = {1\over 2\sin\theta} \left [ \ln {1\over 1-\alpha} + 2\pi ik
-2il\theta\right]\ ,
\end{equation}
where $l,k$ are integers, and $\theta$ runs from $0$ to $\pi$.
Physically, these additional singularities correspond to motion along
periodic orbits of the underlying integrable system (circle), with the
real part ${\rm Re}\ \xi>0$ characterizing the total scattering rate
out of the orbit.

To calculate the level correlation function $R_2^d(\omega)$, we use
the formulas of Section \ref{spectral}, with the spectral function
$S(\omega)$ obtained by substituting the second term of
Eq.~(\ref{mixed11}) in Eq.~(\ref{mixed1}),
\begin{eqnarray}
\label{mixed15}
S(\omega) &=& \sum_l S_l(\omega), \nonumber \\
S_l(\omega) & = & - \left({R\over v_F}\right)^2
{\partial^2\over \partial
z^2} \ln \tilde{J}_l(z)|_{z=i\omega R/v_F} \ , \nonumber \\
\tilde{J}_l(z) &=& - 1 + {\alpha\over 2}
\int_0^\pi d\theta {\sin\theta\over
e^{-2il\theta-2z\sin\theta} - (1-\alpha)}\ .
\end{eqnarray}
In particular, the low-frequency behavior has the form (\ref{lowen})
with the coefficient $A$ given at $\alpha\ll 1$ by
\begin{equation}
\label{mixed16}
A=\left[{256\over 9\pi^2} -6\right]{1\over\alpha^2} \simeq -
{3.12\over\alpha^2} \ .
\end{equation}
As previously, this result is valid until the last term ($1/g_b$
correction) in Eq.~(\ref{lowen}) becomes of order unity. This happens at
a frequency which plays the role of the Thouless energy. For $\alpha \ll
1$ this energy scale is $E_c \sim \alpha v_F/R$, which is the inverse of
the characteristic relaxation time at $\alpha\ll 1$. The $\sigma$-model
only applies when the Thouless energy is much larger than the mean level
spacing, {\em i.e.} for $\alpha \gg g_b^{-1}$. For even lower values
of $\alpha$ the system exhibits integrable behavior.

As usual, at frequencies above the effective Thouless energy $E_c$ the
level statistics are totally different from the RMT predictions.  To
demonstrate this, we calculate the smooth part of the level correlation
function for intermediate frequencies $E_c \ll \omega \ll v_F/R$ using
Altshuler-Shklovskii formula (\ref{as}).  We first notice that the
contributions of all $S_l$ with $l \ne 0$ are suppressed by the small
factor $\omega R/v_F$ as compared to the term $S_0$, and thus can be
neglected. Evaluating the integral in Eq. (\ref{mixed15}) at $l = 0$ and
extracting the leading-order term, we find
\begin{equation} \label{mixed17}
R_2^d (\omega) = -\frac{3\alpha^2}{4\pi^2g_b^2} \left( \frac{v_F}{\omega
R} \right)^4 \left[ \ln \left( \frac{\omega R}{\alpha v_F} \right) + O
(1) \right].
\end{equation}
The result (\ref{mixed17}) matches Eq. (\ref{mixed16}) at $\omega \sim
E_c = \alpha v_F/R$ and drops sharply for higher frequencies.

\section{Summary and Discussion} 
\label{conclus}

In this paper we have used the ballistic $\sigma$-model approach to
investigate statistical properties of energy levels and eigenfunctions of
a circular billiard with diffusive surface scattering. For this simple
model of a chaotic system we calculated explicitly non-universal
deviations of the statistical properties from the random matrix theory.
These non-universal properties are determined by the classical dynamics
and turn out to be very different from the two examples available in the
literature -- diffusive systems and ballistic systems with bulk
$\delta$-correlated disorder. We believe that our results are not
specific for a particular model considered but rather reflect
non-universal features of generic chaotic systems and their
qualitative difference from the corresponding properties of systems
with bulk shot-range impurities. Below we summarize our main
findings.  

The spectral statistics (Section \ref{spectral}) deviate from its RMT
form on a frequency scale set by the inverse flight time (playing the
role of the Thouless energy). At higher energies, the level number
variance saturates and oscillates in agreement with
predictions \cite{Berry} for a generic chaotic system. Surprisingly,
the two-level correlation function shows non-decaying (though weak)
oscillations with the period of the mean level spacing at high
frequencies, producing a $\delta$-like spike in the spectral
form-factor at the Heisenberg time. In Section~\ref{mixed} the
analysis of the spectral statistics is generalized to the case of a
mixed boundary condition when the ``Thouless energy'' is
parametrically smaller than the inverse time of flight. 

In Section \ref{parametric} we presented a thorough study of the
parametric level statistics, with magnetic field playing the role of
the external parameter. We identified all relevant regions in the
frequency-magnetic field plane and calculated the parametric two-level
correlation function and the parametric level number variance in all
of them. In particular, a surprising result was obtained in the region
of high magnetic fields, where the parametric correlation function was
found to be independent of the magnetic field. In other words, the
density of states retain finite memory even in the limit of very large
fields after levels have undergone arbitrary many avoided crossings.

In Section \ref{eigen} we analyzed spatial correlations of eigenfunction
intensities. Since naive application of the ballistic $\sigma$-model led
us to the problem of double counting (with one and the same contribution
appearing twice -- classically and quantum-mechanically), we had to
reanalyze the $\sigma$-model derivation.  For this purpose we introduced
a smooth random potential and obtained the $\sigma$-model from averaging
over it, following Refs.~\onlinecite{woelfle84} (see Appendix). We
have found that while in the limit $\tau/\tau_{tr}\to 0$ (with
$\tau^{-1}$ and $\tau_{tr}^{-1}$ being total and transport scattering
rates, respectively) the obtained action takes the conventional form
(\ref{wf6}) of the ballistic $\sigma$-model, the behavior of the
propagator is different at short distances. This affects the wave
function correlations at short spatial scales $r\lesssim v_F\tau$.
The final result has a form (\ref{wf14}) with the quantum ($k_q$) and
classical ($f_1$) terms corresponding to the motion before and after
the first collision respectively.

In the present context of a system with diffuse boundary scattering,
introduction of an additional smooth random potential
satisfying $d \ll l \ll R \ll l_{tr}$ can be considered as
a technical trick allowing to obtain the $\sigma$-model in the
conventional form (\ref{model1}). In principle, it should be possible
to derive the $\sigma$-model directly by averaging over the
boundary disorder, but in this case the action will be more
complicated, since the system size will essentially play the role of
$v_F\tau$. Let us stress that the additional random potential does not
affect the results for the energy level statistics and for the {\it
smoothed} eigenfunction correlations. On the other hand, the
corresponding mean free path $l=v_F\tau$ manifests itself explicitly
in Eq. (\ref{wf14}) for the eigenfunction correlations by setting the
scale at which the Friedel-type oscillations get smeared.       

We believe that averaging over an additional smooth quantum random
potential is of conceptual importance for the problem of non-universal
features in the level and eigenfunction statistics of a conventional
chaotic billiard (without boundary scattering). Indeed, a consensus
seems to have been reached by now that the energy averaging by itself
is insufficient, and one has to average over some class of systems
(with the same classical dynamics) in order to derive the
$\sigma$-model \cite{aos-rev}. Furthermore, without such an averaging
one cannot detect non-universal features, since the statistics are not
sufficient \cite{prange97}. A smooth quantum random potential with
parameters chosen in such a way that $v_F\tau_{tr}/R\propto
\hbar^{-a}$ with $a>0$ and  $v_F\tau/R\propto \hbar^b$ with $0<b<1$ is
exactly the required type of ensemble averaging. With this averaging,
the ballistic $\sigma$-model can be rigorously derived. Let us
emphasize that the ballistic action obtained in this way will have the
conventional (obtained by gradient expansion) form (\ref{model1}) only
at length scales exceeding $l=v_F\tau$. If one is interested in
eigenfunction correlations at shorter scales, one should avoid the
gradient expansion and use the more general form (\ref{wf5a}),
(\ref{app2}), as explained in Sec.~\ref{eigen} and in Appendix. 

An additional smooth random potential discussed here resembles the one
introduced by Aleiner and Larkin in Refs.~\onlinecite{AL96,AL97}
where the problem of weak localization in ballistic chaotic systems
was considered. There is, however, a conceptual difference: while
Aleiner and Larkin introduced fictitious disorder in order to mimic
diffraction on boundaries of the billiard, we consider a real random
potential, {\it i.e.} an ensemble of systems, averaging over which
allows one to derive the $\sigma$-model, as explained above.

We close the article by mentioning two open issues.

(i) The problem of repetitions \cite{trace1} (which appear not to be
counted properly in the $\sigma$-model approach) still awaits its
resolution. For the model considered in the present article this
problem does not apply since in view of the diffusive nature of
boundary scattering, all directions of motion after a scattering event
are allowed so that the repetitions are irrelevant. 

(ii) We used the boundary condition for diffusive scattering in a linear
form, i.e. we supplemented the Liouville operator determining the
quadratic form of the action by the boundary condition. This was
sufficient for the problems considered in the present article, when the
relevant $\sigma$-model correlation functions are determined by the
structure of the action in the vicinity of spatially homogeneous
configurations and therefore the results are governed by the
eigenfunctions and eigenvalues of the Liouville operator. However, in
general, this is not sufficient, and a boundary condition on the
$g(\bbox{r},\bbox{n})$ field is needed. In particular, one would need
such a general boundary condition to calculate ``tails'' of various
distribution functions (of relaxation times, eigenfunction amplitudes,
local density of states, inverse participation ratio, etc.), analogously
to how it has been done for diffusive systems \cite{als,Mirlin}.

\section*{Acknowledgments}

Useful discussions with I.L.~Aleiner are gratefully acknowledged.
This work was supported by the Swiss National Science Foundation
(Y.~M.~B.), the SFB 195 der  Deutschen Forschungsgemeinschaft
(A.~D.~M.), and the INTAS grant 97-1342 (A.~D.~M.). We also
acknowledge the hospitality of the Lorentz Center, Leiden
(Y.~M.~B. and A.~D.~M.), the Max-Planck-Institut f\"ur Physik
Komplexer Systeme, Dresden (Y.~M.~B. and A.~D.~M.), University of
Geneva (A.~D.~M. and B.~A.~M.), and Forschungszentrum Karlsruhe
(Y.~M.~B.), where parts of this work were carried out.

\appendix
\section{Ballistic $\sigma$-model from small-angle scattering}

Here we outline derivation of the $\sigma$-model based on averaging
over a smooth random potential $U({\bbox r})$ (see
Section~\ref{eigen}), following Ref.~\onlinecite{woelfle84}.
After the averaging and the Hubbard-Stratonovich decoupling by a
supermatrix field ${\cal Q}({\bbox r},{\bbox r'})$ one finds the
action 
\begin{eqnarray}
\label{wf2}
F[\cal Q] &=& {\rm Str}\ln [E+i\frac{\omega}{2}\Lambda-\hat{H}_0
-{\cal Q}({\bbox r},{\bbox r'})V({\bbox r}-{\bbox r'})] \nonumber \\
&+& {1\over 2}{\rm Str}\int d{\bbox r} d{\bbox r'} {\cal Q}({\bbox
r},{\bbox r'}) 
V({\bbox r}-{\bbox r'}){\cal Q}({\bbox r'},{\bbox r})\ ,
\end{eqnarray}
where $\hat{H}_0=\hat{\bbox p}^2/2m$ is the free Hamiltonian.
The corresponding saddle-point equation has the form of the
self-consistent Born approximation (SCBA) and possesses a set of 
translationally-invariant solutions ${\cal Q}({\bbox r},{\bbox r'})
={\cal Q}({\bbox r}-{\bbox r'})$, 
which can be most conveniently written down in the momentum space,
\begin{equation}
\label{wf3}
{\cal Q}({\bbox p}) = {\rm Re} G_R(p) + i T\Lambda T^{-1}{\rm Im}G_R(p)\ .
\end{equation}
Here $G_R(p)$ is the retarded Green's function in SCBA,
\begin{eqnarray}
\label{wf4}
&& G_R(p)=\left(E-H_0(p)+{i\over 2\tau_p}\right)^{-1}\ , \nonumber \\
&& {1\over 2\tau_p} = - \int(d{\bbox p}_1) V({\bbox p}-{\bbox p}_1)
\,{\rm Im}\,G_R(p_1)\ ,
\end{eqnarray}
and the matrices $T$ belong to the super coset space
$U(2|1,1)/U(1|1)\times U(1\vert 1)$. Allowing for slow variation of $T$
with the spatial coordinate ${\bbox r}_+=({\bbox r}+{\bbox r'})/2$ and with
the direction ${\bbox n}$ of the momentum ${\bbox p}$ yields the soft
modes,
\begin{equation}
\label{wf5}
Q({\bbox r}_+,{\bbox n})= T({\bbox r}_+,{\bbox n})\Lambda 
T^{-1}({\bbox r}_+,{\bbox n}) 
\end{equation}
with $Q^2({\bbox r}_+,{\bbox n})=1$. The action for these soft modes has
the form (we set $\omega = 0$)
\begin{eqnarray}
\label{wf5a}
F &=& \mbox{Str}\ln\left(E-\hat{H}_0-{i\over 2}\int d{\bbox n}
Q({\bbox r},{\bbox n'}) w({\bbox n},{\bbox n'})\right)
\nonumber\\
& - & {\pi\nu\over 4} \int d{\bbox r}d{\bbox n}d{\bbox n'} \,{\rm Str}\,
Q({\bbox r},{\bbox n})w({\bbox n},{\bbox n'})Q({\bbox r},{\bbox n'})\ ,
\end{eqnarray}
where $w({\bbox n},{\bbox n'})=2\pi\nu V(p_F|{\bbox n}-{\bbox n'}|)$
is the scattering cross-section. Performing now the gradient expansion
of Eq. (\ref{wf5a}) and using $\tau_{tr}\gg\tau$, one gets the action
of the ballistic $\sigma$-model (generalizing Eq.~(\ref{model1}) to
the case of a non-isotropic disorder scattering), 
\begin{eqnarray}
\label{wf6}
\tilde{F} &=& \pi\nu v_F \int d{\bbox r}d{\bbox n}\,{\rm Str}\, \Lambda 
T^{-1}({\bbox r},{\bbox n}){\bbox n}{\bbox \nabla}T({\bbox r},{\bbox n})
\nonumber\\
& + & {\pi\nu\over 4} \int d{\bbox r}d{\bbox n}d{\bbox n'} \,{\rm Str}\,
Q({\bbox r},{\bbox n})w({\bbox n},{\bbox n'})Q({\bbox r},{\bbox n'})\ .
\end{eqnarray}
Let us emphasize that the action (\ref{wf5a}) takes the form
(\ref{wf6}) only in the limit $\tau/\tau_{tr}\to 0$. While for most
purposes the difference between these two formulas is irrelevant, it
is of crucial importance for  settling the double counting
problem considered in Sec.~\ref{eigen}, since it is related to the
short-scale behavior of the $\sigma$-model propagator.
For this reason we use there
the action  in the form (\ref{wf5a}) (and not the approximation (\ref{wf6})). 

To illustrate the connection and the difference between the actions
(\ref{wf5a}) and (\ref{wf6}), it is instructive to write down
their quadratic forms at low momenta. We write 
$T({\bbox r},{\bbox n}) = 1-W({\bbox r},{\bbox n})/2$ and 
introduce the angular
harmonics of the $W$-field, $W({\bbox r},m)=\int d{\bbox n}
\exp(-im\phi_{\bbox n}) W({\bbox r},{\bbox n})$, and of the scattering
cross-section, $w_m=\int (d\phi/2\pi)e^{im\phi}w(\phi)$. 
The result for the quadratic form of the action (\ref{wf5a}) then
reads \cite{woelfle84}
\begin{equation}
\label{app1}
F_0 = {1\over 2}\int (d{\bbox q}){\rm Str}
W_{-m,12}(-{\bbox q})\Gamma_{mm'}({\bbox q})W_{m',21}({\bbox q})\ ,
\end{equation}
where
\begin{eqnarray}
\label{app2}
\Gamma_{mm'}({\bbox q}) & \simeq & \pi\nu\left[{w_m\over
w_0}(w_0-w_m)\delta_{mm'} +{v_F^2\over 2w_0}q^2\delta_{m0}\delta_{m'0}
\right.\nonumber\\
&-&\left. {iv_F\over 2}{w_mw_{m'}\over w_0^2}
(\bar{q}\delta_{m,m'-1}+\bar{q}^* \delta_{m,m'+1}) \right]\ , 
\end{eqnarray}
where $\bar{q}=q_x+iq_y$. On the other hand, the quadratic terms in the
action (\ref{wf6}) are given by Eq. (\ref{app1}) with the kernel
$\Gamma$ replaced by 
\begin{eqnarray}
\label{app3}
\tilde{\Gamma}_{mm'}({\bbox q})&=&
\pi\nu [(w_0-w_m)\delta_{mm'}\nonumber\\ 
&-&{iv_F\over 2}(\bar{q}\delta_{m,m'-1}+\bar{q}^*
\delta_{m,m'+1})]\ .
\end{eqnarray}
Inverting Eqs.~(\ref{app2}), (\ref{app3}) at small $q$, one finds that
the corresponding propagators ${\cal D}=\Gamma^{-1}$ and 
$\tilde{{\cal D}}=\tilde{\Gamma}^{-1}$ are identical, up to a constant
term,
\begin{equation}
\label{app4}
{\cal D}_{mm'}({\bbox q})-\tilde{{\cal D}}_{mm'}({\bbox q})
={1\over\pi\nu w_m}\delta_{mm'}\ .
\end{equation}
The physical meaning of this 
difference becomes clear from the calculation in Section~\ref{eigen}
(avoiding the momentum expansion). Specifically, the propagator for
the action S is given by a series of ladder diagrams, beginning from
the term ``with $-1$ scattering'' [the term $B^{-1}$ in
Eq.~(\ref{wf9})], while that for the action $\tilde{S}$ starts from
the term with zero collisions (free motion), i.e. from the second term
on the r.h.s. of Eq.~(\ref{wf9}). As a result, $\tilde{{\cal D}}={\cal
D}-B^{-1}$, which reproduces exactly Eq.~(\ref{app4}).

\end{multicols}


\begin{references}
\bibitem{other1} This is not true for ``tails'' of distribution
functions of various quantities characterizing wave functions, which
are determined by anomalously localized states (see
Section~\ref{conclus} for discussion). We do not consider such tails
in the present article.
\bibitem{lastfoot} Note that our calculations are in fact limited by
the condition $\omega\ll E_F$ leading to a broadening of the
$\delta$-spike with a width $\sim 1/g_b^2\Delta$.
\bibitem{foot1} For weak disorder, the boundary scattering is almost
specular (Eq. (\protect\ref{bound3}) with $\alpha \ll 1$). For very
strong disorder the boundary sites are effectively decoupled from the
rest of the system, and the scattering is again almost specular.
\end{references}
\end{document}